\documentclass[trackchanges,tighten,twocolumn,twocolappendix,astrosymb]{aastex701}

\usepackage{gensymb}
\usepackage{amsmath,bm}
\usepackage{subcaption}
\usepackage{float}
\usepackage{caption}

\shorttitle{Mixed modes in hydrodynamical simulations}
\shortauthors{De Vries et al.}
\received{XXXX, XXXX}
\revised{YYYY, YYYY}
\accepted{ZZZZ, ZZZZ}
\submitjournal{ApJ}

\begin{document}

\title{Revealing mixed modes in compressible hydrodynamical simulations of red giant stars}

\author[orcid=0000-0002-6272-9839]{Nils B. de Vries}
\affiliation{Department of Physics and Astronomy, University of Exeter, Exeter EX4 4QL, UK}
\email[show]{N.B.de-Vries@exeter.ac.uk}  

\author[orcid=0000-0002-6686-7956]{Arthur Le Saux}
\affiliation{Université Paris-Saclay, Université Paris Cité, CEA, CNRS, AIM, F-91191 Gif-sur-Yvette, France}
\email{Arthur.LESAUX@cea.fr}

\author[]{Isabelle Baraffe}
\affiliation{Department of Physics and Astronomy, University of Exeter, Exeter EX4 4QL, UK}
\affiliation{Ecole Normale Supérieure, Université de Lyon, CRAL (UMR CNRS 5574), F-69007 Lyon, France}
\email{I.Baraffe@exeter.ac.uk}

\author[orcid=0000-0002-0271-5953]{Thomas Guillet}
\affiliation{Department of Physics and Astronomy, University of Exeter, Exeter EX4 4QL, UK}
\email{T.A.Guillet@exeter.ac.uk}

\author[orcid=0000-0002-2522-8605]{Richard H. D. Townsend} 
\affiliation{Department of Astronomy, University of Wisconsin-Madison, 475 N Charter Street, Madison, WI 53706, USA}
\email{townsend@astro.wisc.edu}

\author[]{Armand Leclerc}
\affiliation{Institute of Science and Technology Austria (ISTA), Am Campus 1, 3400 Klosterneuburg, Austria}
\email{Armand.Leclerc@ist.ac.at}

\author[orcid=0000-0002-8261-9143]{Adrien Morison}
\affiliation{Research Software Engineering, MVLS SRF, University of Glasgow, Glasgow}
\affiliation{Department of Physics and Astronomy, University of Exeter, Exeter EX4 4QL, UK}
\email{Adrien.Morison@glasgow.ac.uk}

\begin{abstract}

Mixed modes are observed in many low-mass evolved stars. They provide information about core rotation rates of these stars, which are lower than predicted by stellar evolution models. The mixed modes themselves have been invoked as an angular momentum transport mechanism, but estimating their transport efficiency requires knowledge of their amplitudes. We constrain, for the first time, the mixed mode amplitudes in 2D hydrodynamical simulations of a $1.3M_\odot$ red giant using the code \textsc{music}. We perform two simulations with outer radial truncations at fractional radii $r_o/r_\star = 0.90$ and $r_o/r_\star = 0.98$. We compare the modes in the simulation with those found using both \textsc{gyre} and a \textsc{dedalus} eigenvalue solver. Excellent frequency agreement is found for all p-dominated modes, with minor discrepancies for g-dominated modes, especially in the frequency range $[60, \ 240]\ \mu\mathrm{Hz}$. We find excellent eigenfunction agreement for all modes except those in this frequency range. According to empirical predictions the largest kinetic energies are located around $\nu_{\mathrm{max}} = 312.8\ \mu\mathrm{Hz}$, but in both simulations the modes with frequencies $\nu <50\ \mu\mathrm{Hz}$ have the largest kinetic energies.
In the simulation with $r/r_\star = 0.98$, the simulated modes have extrapolated surface velocities comparable to the empirical predictions, with highest surface velocities in a bell-shaped curve peaking around $\nu = 700 \ \mu\mathrm{Hz}$. The extrapolated surface velocities of the low frequency modes are small, and thus hard to observe, but their large kinetic energies deeper in the interior could significantly impact angular momentum transport, which has not yet been investigated.
\end{abstract}



\section{Introduction}
The CoRoT \citep{Auvergne2009}, Kepler \citep{Borucki2010} and TESS \citep{Ricker2014} missions have revealed the presence of a large amount of stochastically excited acoustic and mixed modes in low-mass subgiant and red giant stars. These mixed modes are standing waves that behave like gravity modes in the core and acoustic modes in the envelope \citep[e.g.][]{Dziembowski1971,Unno1989}. The sensitivity of the mixed modes to conditions in the core sheds light on the structure of stellar interiors \citep[e.g.][]{Bedding2010,Beck2011,Mosser2011}, constraining the core rotation rates of many low-mass subgiant and red giant stars. These stars were found to have a rapidly rotating core and a slowly rotating envelope \citep[e.g.][]{Mosser2012}. This radial differential rotation evolves as the star ages \citep[e.g. ][]{Deheuvels2014,Deheuvels2020,Li2024} and is connected to one of the key open questions of stellar evolution: the role of angular momentum (AM) transport \citep[see for example the review by][]{Aerts2019}. The differential rotation in these stars is much smaller than one would expect from angular momentum conservation alone. Moreover, stellar evolution models incorporating known AM transport mechanisms---such as shear, meridional circulation and magnetic fields generated by the Tayler-Spruit dynamo---also predict negligible AM transport for low-mass evolved stars, yielding core rotation rates 2 orders of magnitude larger than observed \citep{Eggenberger2012,Marques2013,Cantiello2014}. Additional AM transport mechanisms are therefore required.
In order to match the observed ratios of core to envelope rotation rate,
the transport efficiency of these mechanisms is expected to decrease as the star evolves along the subgiant branch \citep{Eggenberger2019a},
before climbing again along the red giant branch \citep{Eggenberger2017}. Computations of AM transport using semi-analytical predictions based on 1D models find that IGWs, excited by convective motions and penetrating plumes \citep[e.g.][]{Press1981,Zahn1997,Lecoanet2013,Pincon2016}, transport AM efficiently on the subgiant branch, with decreasing efficiency as the star evolves \citep[e.g.][]{Fuller2014,Pincon2017}. The observed core rotation rate on the red giant branch has been explained using numerous mechanisms: the revised \citep{Fuller2019} version of the Tayler-Spruit Dynamo \citep{Spruit2002}, the azimuthal magneto-rotational instability (AMRI) \citep[e.g.][]{Spada2016,Moyano2023} and the magnetic webs left behind by the convective core of stars with $M \geq 1.3M_\odot$ \citep{Skoutnev2025}. The mixed modes themselves have also been invoked to transport AM \citep{Belkacem2015a}. Using this semi-empirical formalism, \citet[][]{Belkacem2015b} estimated that the mixed modes do not efficiently transport AM in subgiants and young red giants but do so in more evolved red giants, becoming more efficient as the star evolves. This estimate, however, depends strongly on the amplitudes of the mixed modes. For evolved red giants, observations do not sufficiently constrain mixed mode amplitudes. Therefore, it is uncertain whether mixed modes can transport enough AM to match the observed core rotation rates \citep{Bordadagua2025}. 

We build on a vast array of work performing hydrodynamical simulations to study the amplitudes and excitation of IGWs and g-modes \citep[e.g.][]{Rogers2013,Alvan2014,Edelmann2019,Horst2020,Lecoanet2021,Breton2022,LeSaux2022,Anders2023,Thompson2024, Blouin2023,Blouin2026} and p-modes \citep[e.g.][]{Nordlund2001,Stein2001,Samadi2003,Zhou2019, LeSaux2026}. In particular, \citet{Blouin2023,Blouin2026} performed compressible hydrodynamical simulations of red giants at the bump and tip of the red giant branch. To our knowledge, mixed modes have, however, only been able to be identified in simulations of the carbon and oxygen burning shell of a massive star \citep{Meakin2006} and the Sun \citep{LeSaux2025}.\\
In this work, we study the amplitudes of mixed modes in evolved low-mass stars, which are needed to constrain the contribution of mixed modes to angular momentum transport. 
Therefore, we have, for the first time, studied mixed modes in 2D hydrodynamical simulations of the interior of evolved low-mass stars using the code \textsc{music} \citep[e.g.][]{Viallet2016,Goffrey2017,Baraffe2023}.\\
In Sec.~\ref{sec:model setup} we detail the setup of the simulations and their initial conditions. In Sec.~\ref{sec:Results}, we compare the frequencies and eigenfunctions of the modes in our simulations to predictions from linear theory. In Sec.~\ref{sec:surface_amps} we present the kinetic energies and extrapolated surface velocities of the modes in our simulations. We compare the velocities with the semi-empirical predictions \citep{Belkacem2015b,Bordadagua2025} used for angular momentum transport. Finally, we conclude and discuss the directions of future work with these simulations in Sec.~\ref{sec:conclusions}.

 \section{Model setup and numerical simulations}
 \label{sec:model setup}
 
\begin{table*}[]
\centering
\caption{Parameters of the simulations. $r_i/r_\star$ and $r_o/r_\star$ are the inner and outer fractional radii of the model. $r_\mathrm{conv}/r_\star$ is the fractional radius of the convective boundary. $N_r$ and $N_\theta$ are the grid cells in the $r$ (radial) and $\theta$ (colatitudinal) directions. $\tau_c$ is the convective turnover time, defined in Eq.~\eqref{eq:tau_conv}. $T_\mathrm{sim}$ and $T_\mathrm{steady}$ are the total simulation time and time that the model is in steady state. Finally, $\Delta\nu$ and $\Delta\Pi_{1.797}$ are the large frequency separation and period spacing in the truncated models.}
\centering

\begin{tabular}{llllllllll}
\hline
Model   & $r_i/r_\star$ & $r_o/r_\star$ & $r_\mathrm{conv}/r_\star$ & $N_r \times N_\theta$  & $\tau_\mathrm{c}\ (\mathrm{s})$ & $T_\mathrm{sim}\ (\mathrm{s})$& $T_\mathrm{steady}\ (\mathrm{s})$ & $\Delta\nu\ (\mu\mathrm{Hz})$ & $\Delta \Pi_{1.797}\ (\mathrm{s})$\\ \hline
\textit{RGext90} & 0.02     &  0.90 & 0.214        & $1264 \times 768$ &  $2.9\cdot10^6$      &  $6.63 \cdot 10^7$ &  $2.87 \cdot 10^7$ &  37.74     &  1361.83  \\
\textit{RGext98} & 0.02     &  0.98 & 0.214        & $1392 \times 768$ & $2.0\cdot10^6$  & $6.74 \cdot 10^7$ &  $3.29 \cdot 10^7$  & 27.53 &  1361.42\\ \hline
\end{tabular}
\label{Tab: model_parameters}
\end{table*}

 \begin{figure*}
\subfloat[\footnotesize{Snapshots of the \textit{RGext90} simulation.} \label{fig:snapshot_RGext90}]{\includegraphics[width=\textwidth]{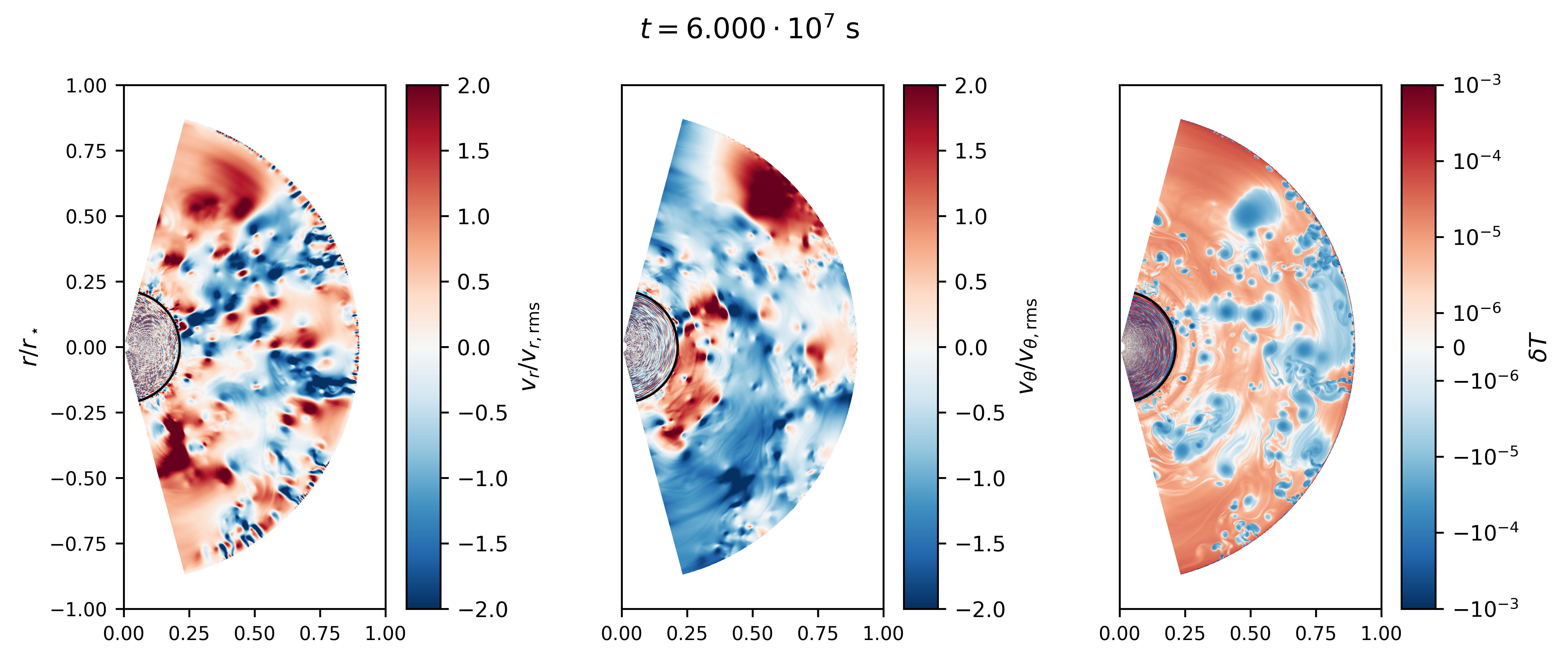}}

\subfloat[\footnotesize{Snapshots of the \textit{RGext98} simulation.}\label{fig:snapshot_RGext98}]{\includegraphics[width=\textwidth]{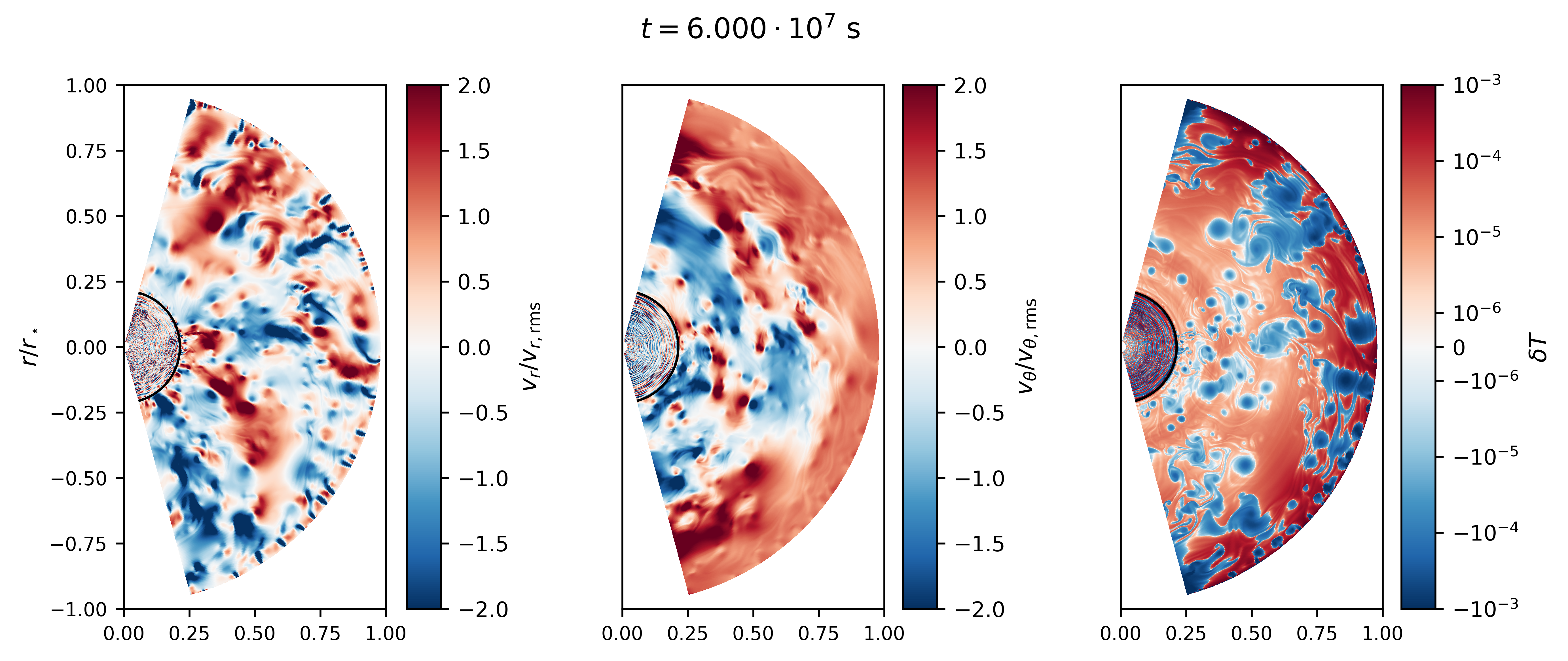}}

\caption{Snapshots of the radial (left) and meridional (middle) velocities normalised by the respective rms velocity at every radius and the temperature perturbation (right) compared to the meridional average temperature. The velocities and temperature perturbations are larger in the \textit{RGext98} simulation compared to the \textit{RGext90} simulation throughout the simulation domain.}
\label{fig:snapshots}

\end{figure*}

We perform 2D hydrodynamical simulations using version 7.0.0 of the fully compressible time-implicit code \textsc{music}, as described in \cite{Viallet2011,Viallet2016,Goffrey2017,Baraffe2023}. We solve the fully compressible non-rotating Euler equations in spherical coordinates with thermal diffusion. We simulate a meridional wedge slice---with meridional extent from $\theta_i = 15\degree$ to $\theta_o = 165\degree$---of the M1 model of \citet{Belkacem2015b}, representing a young $1.3M_\odot$ red giant star. More details of the equations, \textsc{music} itself and the initial stellar model can be found in Appendix~\ref{app:MUSIC_code}.
We compare two simulations of this stellar model with different outer radial truncations to also understand the effect of the upper convective region on the mode amplitudes. Both simulations have an inner radial boundary of $r_i/r_\star = 0.02$, where $r_\star  = 2.586\cdot10^{11}\ \mathrm{cm}$ is the photospheric radius of the stellar model. The simulation labelled  \textit{RGext90} has an outer boundary $r_o/r_\star = 0.90$ and the simulation labelled \textit{RGext98} has $r_o/r_\star = 0.98$. %
Both these configurations include a large convective region to stochastically excite the (mixed) modes, as well as a large part of the stably stratified region to house the g-mode part of the mixed modes. Increasing the outer truncation radius in these 2D stellar models is expected to increase the flow velocities \citep[e.g.][]{Vlaykov2022}; this is expected to result in waves and modes with larger amplitudes in \textit{RGext98}. Indeed, the analysis of amplitudes of IGWs in \citet{Blouin2023} found that increasing the radial truncation increases the flow velocities which in turn increase the amplitudes of the IGWs.

Each simulation is run starting from the initial conditions until the convection, waves and modes are established. At this point the simulation is considered to be in dynamical steady-state. It is then run for another time period $T_\mathrm{steady}$,   during part of this time the spectrum is captured.
Parameters of both these simulations are shown in Table~\ref{Tab: model_parameters}, where $\tau_c$ is the convective turnover time given by 
\begin{equation}
    \tau_c = \left\langle\int_{r_i}^{r_o}\frac{1}{\sqrt{\langle\bm{v}\cdot\bm{v}\rangle_\theta}} \ \! \mathrm{d}r\right\rangle_t,
    \label{eq:tau_conv}
\end{equation}
with $\bm{v}$ the velocity, $\langle\cdot\rangle_\theta$ the meridional average and $\langle\cdot\rangle_t$ the time average computed over the steady state. In addition to simulation parameters we also report two illustrative asteroseismic quantities computed from the truncated models in this table: the large frequency separation, $\Delta\nu$, which represents the frequency between two adjacent p-modes and the period spacing $\Delta\Pi_{1.797}$, which represents the period spacing between two adjacent g-modes with $\ell = 1.797$. Non-integer values of $\ell$ result from the employed wedge harmonics, see Appendix~\ref{app:MUSIC_code} for more details.

\section{Analysis of the flow in the simulations}
\label{app:flow_snapshots}

 \begin{figure*}
 \centering
\subfloat[\footnotesize{The \'Echelle diagram at the radius of $r/r_\star = 0.89$.} \label{fig:Echelle_spectrum_r_0.89}]{
         \includegraphics[width=0.45\textwidth,trim={0.5cm 0 1cm 1.2cm},clip=true]{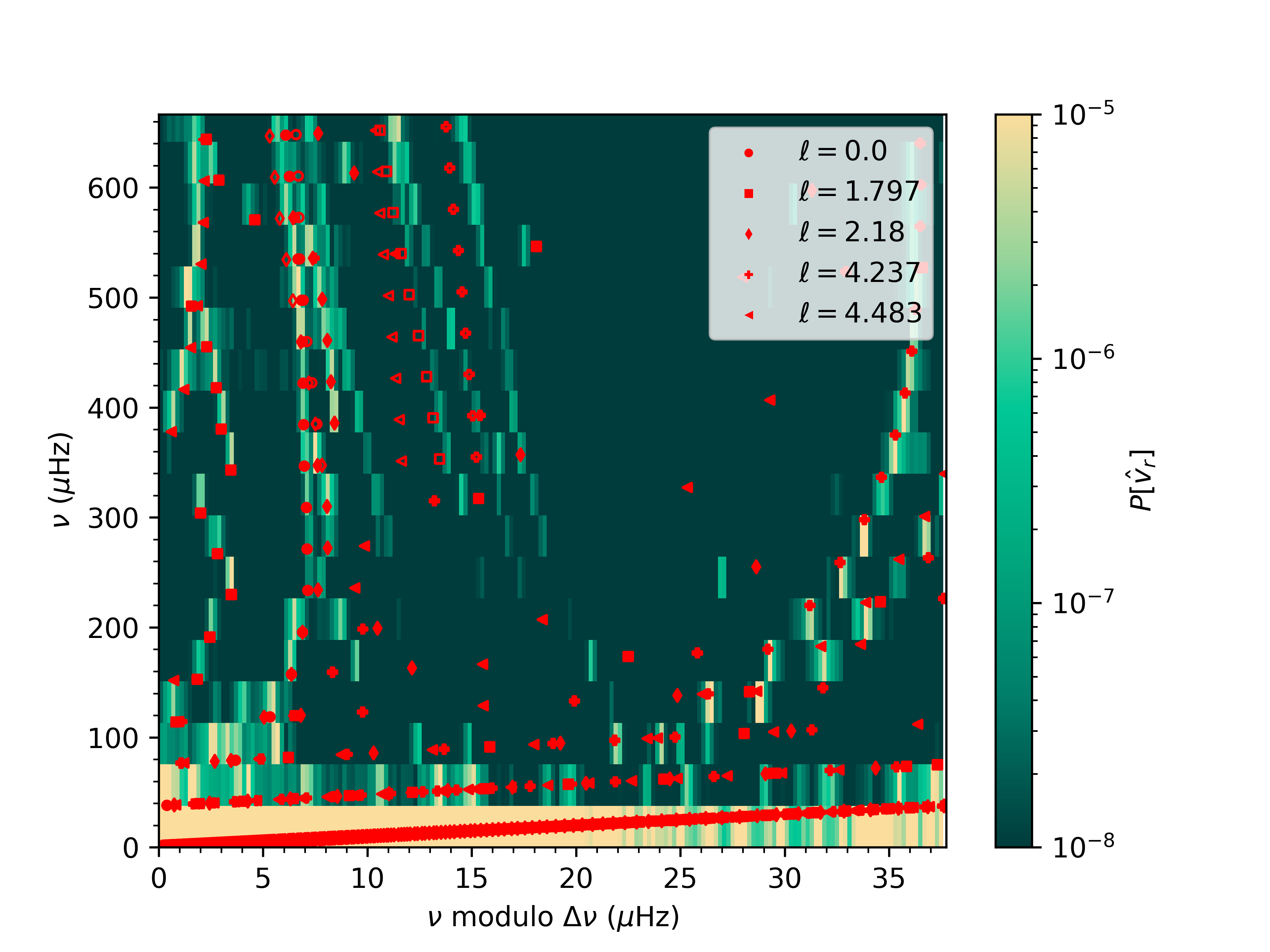}}
\hspace{0.06\textwidth}
\subfloat[\footnotesize{The frequency-radius diagram with $\ell = 0$.} \label{fig:freq_rad_spectrum_l_0.0}]{\includegraphics[width=0.45\textwidth,trim={0.5cm 0 1cm 1.2cm},clip=true]{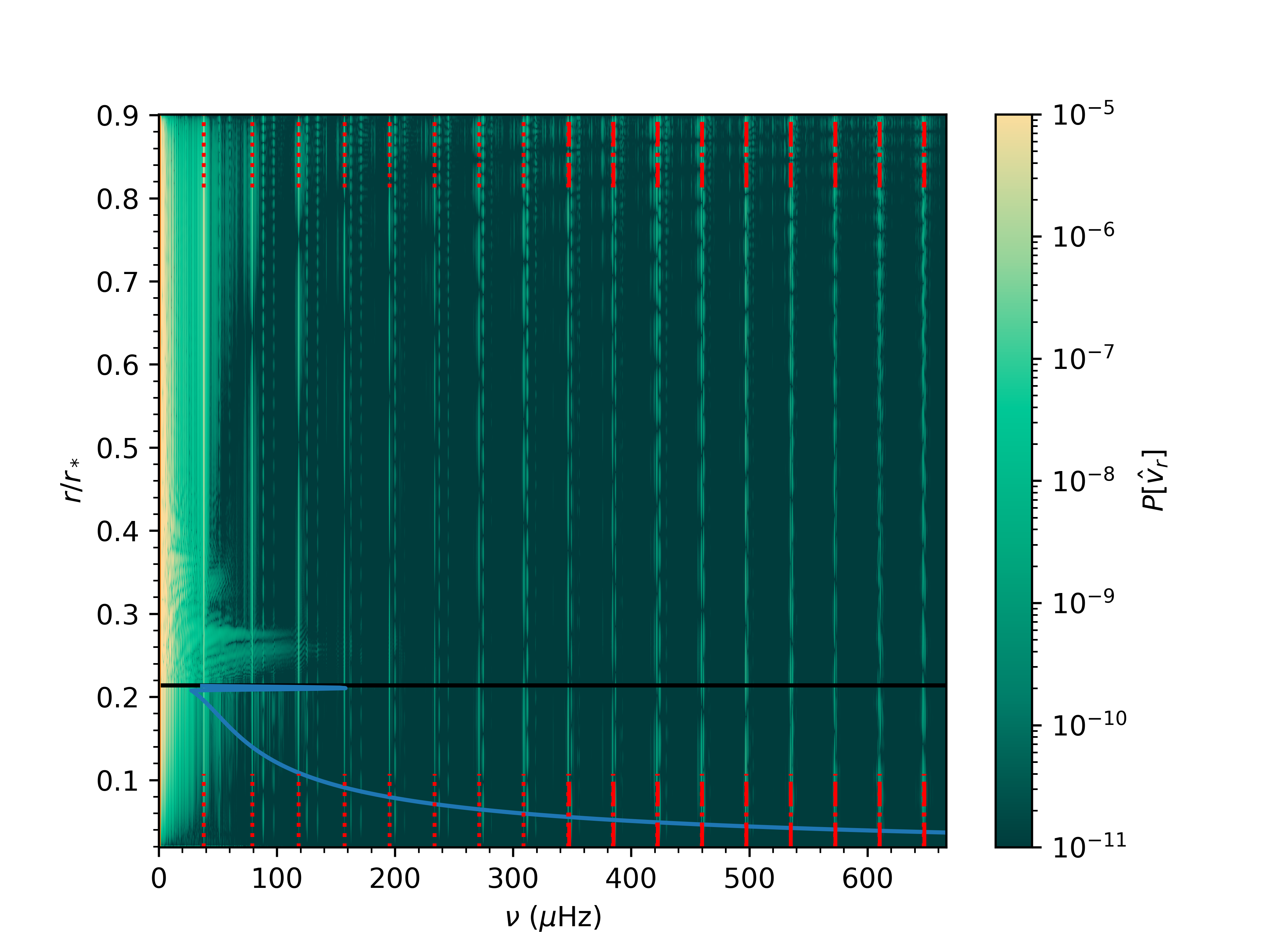}}\\

\centering
\subfloat[\footnotesize{The frequency-radius diagram with $\ell = 1.797$.}\label{fig:freq_rad_spectrum_l_1.797}]{
         \includegraphics[width=0.45\textwidth,trim={0.5cm 0 1cm 1.2cm},clip=true]{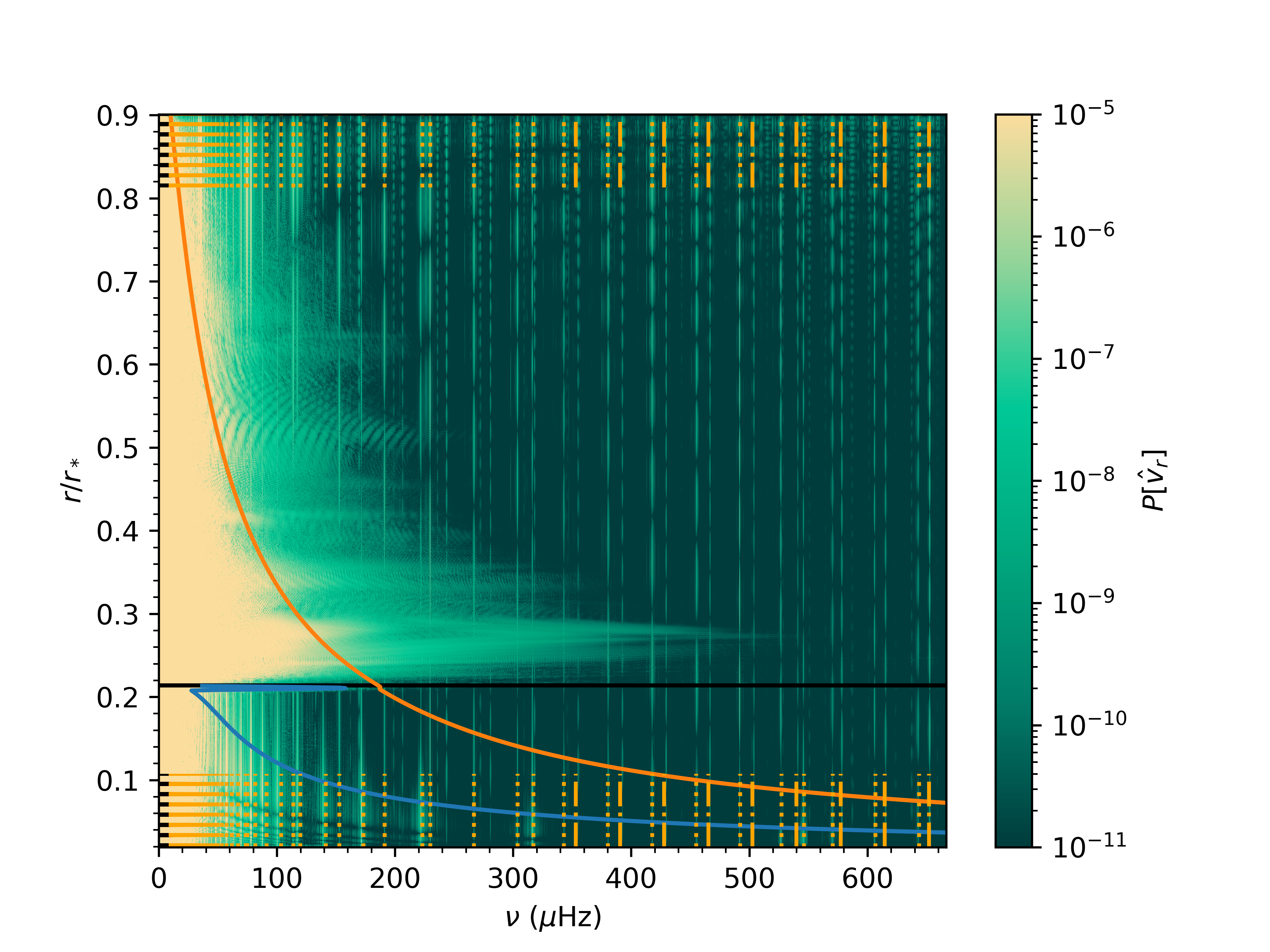}}
\hspace{0.06\textwidth}
\subfloat[\footnotesize{The frequency-radius diagram with $\ell = 4.237$.} \label{fig:freq_rad_spectrum_l_4.237}]{\includegraphics[width=0.45\textwidth,trim={0.5cm 0 1cm 1.2cm},clip=true]{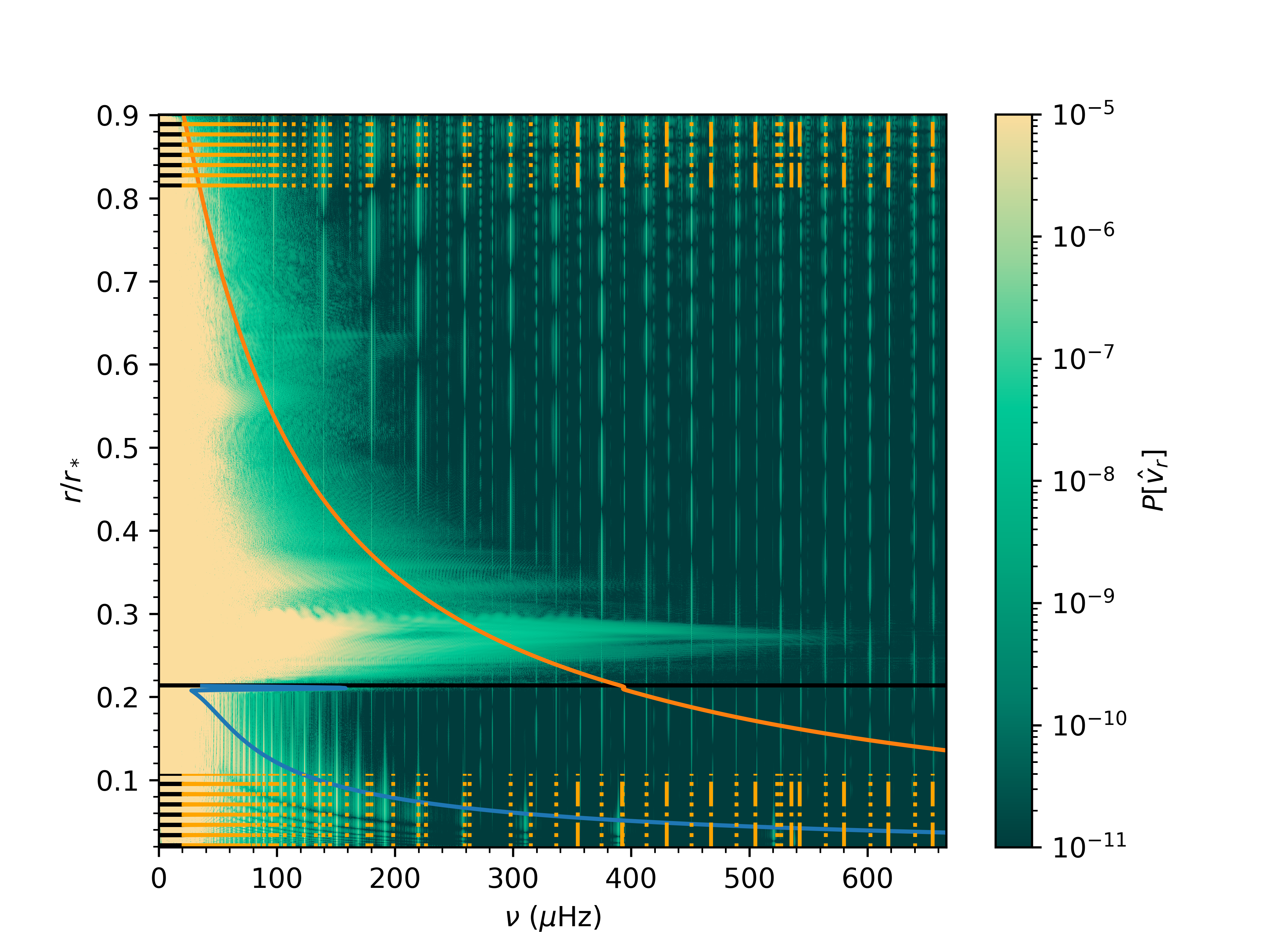}}\\

     \caption{Spectra of $P[\hat{v}_r](\nu, r, \ell)$. In panel a) we show an \'Echelle diagram, with the linear modes overplotted in red. Filled symbols indicate modes located at that frequency, while spectral reflections are plotted in open symbols. In the other three panels we show frequency-radius spectra at fixed values of $\ell$, note the different color scales for these. In the frequency-radius diagrams we plot the buoyancy frequency $N$, the lamb frequency $S_\ell$ and convective boundary in solid-blue, solid-orange and solid-black respectively. The frequencies predicted by the linear theory for modes with only p-mode character (red), only g-mode character (black) and both characters (orange) are plotted in vertical lines. Dotted lines represent the modes at that frequency, while dash-dotted lines represent spectrally-folded modes.}
    \label{fig:spectra_main}

\end{figure*}

In Fig.~\ref{fig:snapshots} we show snapshots of the simulations \textit{RGext90} and \textit{RGext98} taken at time $t=6\cdot10^7 \ \mathrm{s}$, within the period over which the spectrum for this simulation is computed. In the left and middle panels of this figure the radial and meridional velocity perturbations are plotted. Both are normalised by their rms velocity at each radius, defined by:

\begin{equation}
    v_{i,\mathrm{rms}}(r) = \sqrt{\langle v_i^2\rangle_\theta},
\end{equation}
where $i = r,\theta$. The temperature deviation from the meridional average of the temperature of this snapshot

\begin{equation}
    \delta T = \frac{T- \langle T\rangle_\theta}{\langle T\rangle_\theta},
\end{equation}

 is plotted in the right panels. The convective boundary as specified by the 1D model is shown in the solid-black line. The radial velocity in the left-hand panel of Fig.~\ref{fig:snapshot_RGext90} shows large scale motions that reach the convective boundary and excite IGWs there. These IGWs, with large radial wavenumbers and thus short radial wavelengths, form standing modes that appear as spirals in the radiative region in the inner part of the simulation. The meridional velocity in the middle panel shows a large scale clockwise flow, plotted in blue, has developed throughout the convective region of the simulation. This meridional flow is enabled by the periodic boundary conditions. Finally, the temperature perturbation is shown in the right-hand panel. Comparing this simulation to the simulation \textit{RGext98} in Fig.~\ref{fig:snapshot_RGext98}, we find that the latter has larger flow velocities throughout the domain due to the large velocities at the top, as expected from increasing the outer truncation radius towards the surface \citep[e.g.][]{Vlaykov2022,Blouin2023}. The temperature perturbations are also significantly larger in the simulation \textit{RGext98}, as also found in \citet{Vlaykov2022}. Qualitatively, the larger radial extent has not significantly modified the look of the radial velocity in the panel of the left. In the meridional velocity a fast counterclockwise large scale meridional velocity has developed in the outer edge of the simulation domain. We expect the impact of the convective meridional flow on mixed modes to be small: in the convective region, mixed modes behave as p-modes and are therefore mostly radial waves, and neither the pressure nor radial velocity profiles seem to be significantly impacted by the meridional circulation. The effects of the meridional velocity on the power spectrum of convection and thus its indirect effect on mode excitation requires further study by completing simulations with free-slip boundary conditions in the horizontal direction, which should prevent the meridional circulation.

\section{Agreement between linear theory and non-linear simulations}
\label{sec:Results}

\subsection{Comparison of the mode frequencies}
In this section we study the agreement between the frequencies predicted by linear theory and those retrieved from the 2D non-linear simulations. High levels of agreement show that the simulations capture linear theory well, and that we can safely determine surface velocities by extrapolating the modes using linear theory. We find this agreement to be similar for both simulations, so we only show results for the simulation \textit{RGext90}. We will compare the two simulations again when we examine the kinetic energies and surface velocities of these modes in Sec.~\ref{sec:surface_amps}. To study the modes that establish in our simulation and compare to the predictions from linear theory, we take the radial velocity spectra obtained from these simulations, $P[\hat{v}_r](\nu, r, \ell)$. The relation between the power spectrum and velocity is given by $P[\hat{v}_r](\nu, r, \ell) = 2|\hat v_r(\nu,r,\ell)|^2$. The spectra are obtained by first taking the wedge harmonics transform, followed by the Fourier transform, of the radial velocity field of the simulations; more details on this method can be found in e.g.~\citet{LeSaux2022,LeSaux2025}. We use the radial velocity because this is the velocity component with lowest convective signal in the simulations. 

 We first construct an \'Echelle diagram of all spectra summed up to $\ell < 5$ in Fig.~\ref{fig:Echelle_spectrum_r_0.89} to study the agreement between the frequencies predicted by linear theory to the modes we find in the simulations. We only analyse the spectra with $\ell <5$ because we expect no significant coupling of mixed modes above this value of $\ell$. We choose to use the fractional radius of $r/r_\star = 0.89$ for the \'Echelle diagram, just below the upper boundary of this simulation, as it shows the p-mode parts of the modes, thus most resembling the \'Echelle diagrams obtained from observations. We also note the strong convective signal at low frequencies, which makes it hard to identify modes with such frequencies. We compare the results to linear theory using both version 8.0.1 of \textsc{gyre}, with additionally implemented non-integer values of $\ell$, as well as the option to set $v_r=0$ outer boundary conditions for $\ell = 0$, to match our simulations, as well as an eigenvalue solver written in \textsc{dedalus} \citep{Burns2020} based on the one employed in \citet{Perrot2019,Leclerc2022,Leclerc2024,LeSaux2025}. More details on this eigenvalue solver and its setup as well as the settings used for \textsc{gyre}  can be found in Appendix~\ref{app:Dedalus_code}. These linear calculations are based on the 1D initial model. We use both linear solvers as a check for robustness against each other. We find excellent agreement between them for both frequencies and mode eigenfunctions, as shown in Appendix~\ref{app:Ded_GYRE_comparison}. Only the results of the \textsc{dedalus} eigenvalue solver are plotted in Fig.~\ref{fig:spectra_main}. We overplot the frequencies predicted by linear modes in red markers on the \'Echelle diagram in Fig.~\ref{fig:Echelle_spectrum_r_0.89}. The different values of $\ell$ are represented by different symbols. 
 The sampling rate of the spectra is $1333.33\  \mu\mathrm{Hz}$, resulting in a Nyquist frequency of $666.67\  \mu\mathrm{Hz}$. Filled symbols indicate modes that are located at that frequency, while spectral reflections from modes with frequencies above the Nyquist frequency, but below $1000\ \mu\mathrm{Hz}$ are plotted in open symbols. We find good agreement between the modes at $\ell = 0,\ 1.797,\ 2.18$ obtained from linear theory and those found in the simulation. In particular we can see the near-vertical ridge of the radial $\ell = 0 $ modes at high frequency. In addition we can see clear avoided crossings in the $\ell = 1.797$ and $\ell = 2.17$ curves. The agreement of linear theory and the modes in the simulation that lie closest to these nearly vertical branches, i.e. the p-dominated modes, is excellent below the Nyquist frequency. The agreement becomes slightly worse as the frequency gets closer to the Nyquist frequency, but is still $\lesssim1\ \mu\mathrm{Hz}$. A clear discrepancy is visible in the spectrally reflected modes with frequencies larger than the Nyquist frequency. For the curves at $\ell = 4.237,\ 4.483$ we note that the agreement is also very good for the modes on the main branches. These main branches seem to take a similar shape to the main branches at lower $\ell$, but are more curved, only tending to a near-vertical ridge for $\nu \gtrsim 450\ \mu\mathrm{Hz}$ upwards. The modes away from all these main branches, i.e. the g-dominated modes, for $\ell =1.797,\ 2.18$ show disagreement of a few $\mu\mathrm{Hz}$, particularly for modes with $\nu \lesssim 240\ \mu\mathrm{Hz}$. The g-dominated modes for $\ell = 4.237,\ 4.483$ do not emerge at this radius because the amplitude of the p-mode part of these modes is small. Examining the same diagram at $r/r_\star = 0.03$ (not shown) we find that such g-dominated modes with $\ell = 4.237,\ 4.483$ show similar or larger frequency discrepancies than the ones with $\ell =1.797,\ 2.18$.\\

 \begin{figure*}
 \centering
\subfloat[\footnotesize{A pure p-mode at $\ell = 0$, $\nu = 195.5\ \mu\mathrm{Hz}$} \label{fig:modes_pure_p_l_0.0}]{
         \includegraphics[width=0.42\textwidth,trim={0.1cm 0.1cm 0.3cm 1.2cm},clip=true]{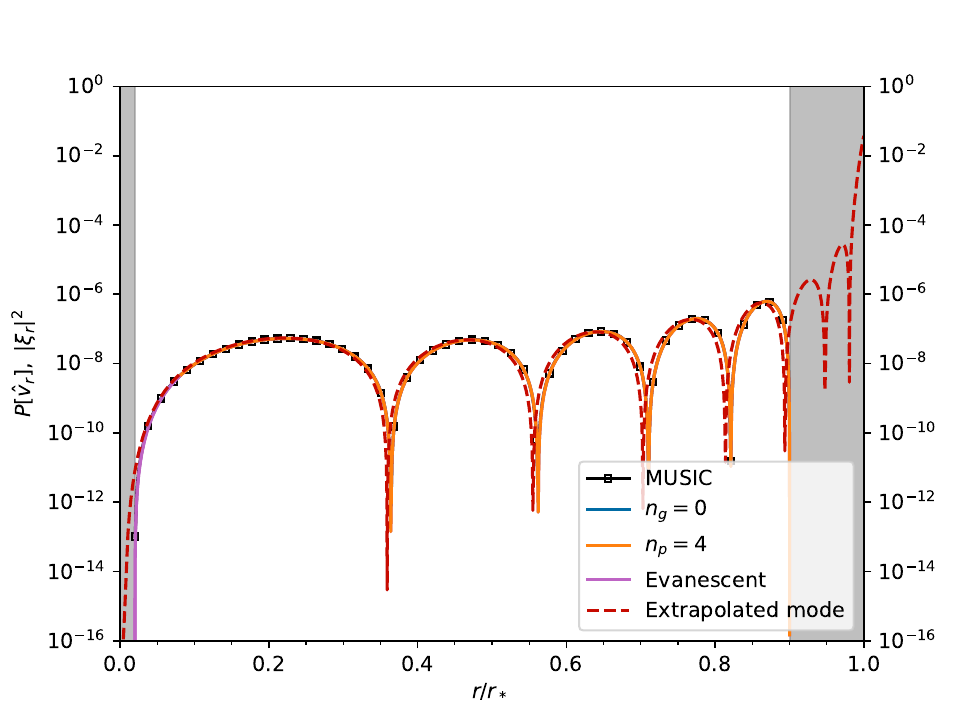}}
\hspace{0.06\textwidth}
\subfloat[\footnotesize{A f/g mixed mode at $\ell = 1.797$, $\nu = 36.0\ \mu\mathrm{Hz}$} \label{fig:modes_mixed_low_freq_l_1.797}]{\includegraphics[width=0.42\textwidth,trim={0.1cm 0.1cm 0.3cm 1.2cm},clip=true]{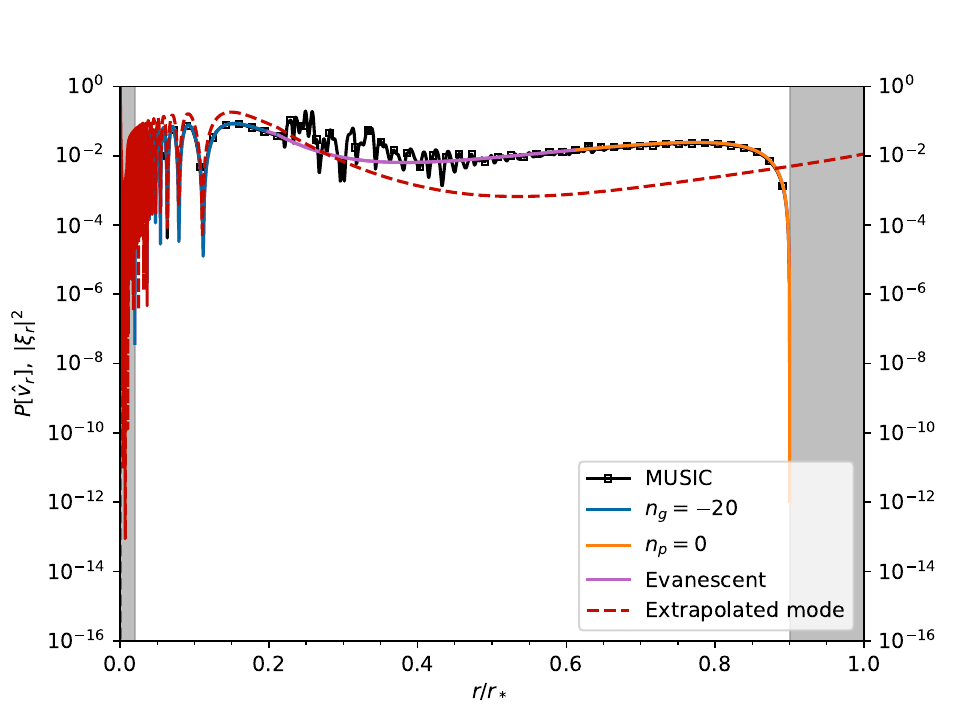}}\\

\centering
\subfloat[\footnotesize{A p-dominated mixed mode at $\ell = 2.18$, $\nu = 460.7\ \mu\mathrm{Hz}$} \label{fig:modes_mixed_high_freq_l_2.18}]{
         \includegraphics[width=0.42\textwidth,trim={0.1cm 0.1cm 0.3cm 1.2cm},clip=true]{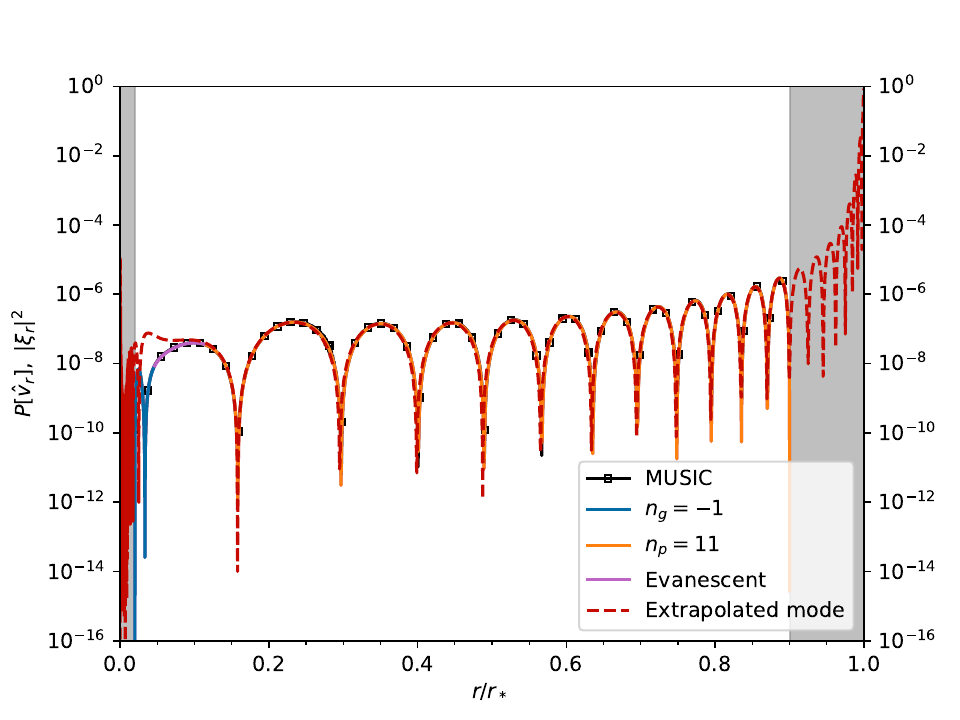}}
\hspace{0.06\textwidth}
\subfloat[\footnotesize{An ill-matching mixed mode at $\ell = 1.797$, $\nu = 113.8\ \mu\mathrm{Hz}$} \label{fig:modes_mixed_bad_fit_l_1.797}]{\includegraphics[width=0.42\textwidth,trim={0.1cm 0.1cm 0.3cm 1.2cm},clip=true]{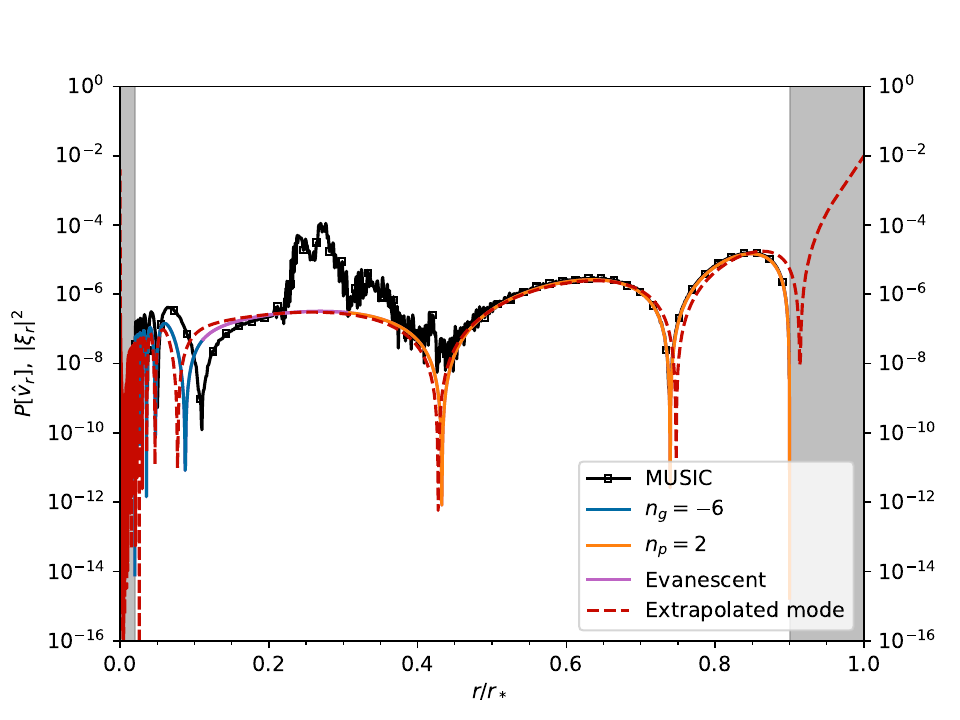}}\\

     \caption{The simulation spectrum $P[\hat{v}_r](\nu, r, \ell)$ at the chosen $\ell$ and frequency is plotted as a function of radius in solid-black with square black markers. This signal includes both the mode and the convective signal. On top of the simulation results the radial displacement eigenfunctions obtained from linear theory are plotted in blue, orange and purple, corresponding to regions where the mode acts as a g-mode with $n_g$, p-mode with $n_p$ and is evanescent respectively. These linear theory results lie on top of the simulation results due to their excellent agreement, almost completely obscuring the simulation results except for the black square markers. The region that is outside the simulation domain is shown in light grey. The extrapolation of the modes to the full star is plotted in dashed-red.}
    \label{fig:modes_main}

\end{figure*}

Following e.g. \citet{Alvan2014,Edelmann2019} we study the modes present in our simulations using frequency-radius diagrams at fixed values of $\ell$, plotted in the other three panels of Fig.~\ref{fig:spectra_main}. The buoyancy frequency $N$ and the lamb frequency $S_\ell$ are plotted on these figures in solid-blue and solid-orange respectively, while the convective boundary is plotted in solid-black. Note the different color scale in these panels compared to Fig.~\ref{fig:Echelle_spectrum_r_0.89}, necessary to clearly show the modes in the bulk of the simulation domain, where the radial velocity is expected to be smaller than near the top. The frequencies of the modes found using the \textsc{dedalus} eigenvalue solver are plotted using vertical dotted lines at the top and bottom of this figure.  
Vertical dash-dotted lines represent spectrally folded modes. More modes than just the identified modes are visible in the spectra as there are more spectrally folded modes than the frequencies we have calculated. The spectrum at $\ell = 0$ is plotted in Fig.~\ref{fig:freq_rad_spectrum_l_0.0} showing excellent agreement between simulated and linearly predicted modes. The spectrally-folded modes and regular modes at frequencies $\gtrsim 350 \ \mu\mathrm{Hz}$ overlap by coincidence. Only p-modes arise because g-modes are prohibited at $\ell = 0$. A continuum of signal is present a low frequencies, corresponding to convective signal. A visual examination of the modes reveals that the mode with the largest velocities throughout the domain in this simulation is the lowest frequency p-mode. The high frequency modes appear much weaker than this lowest frequency p-mode. 
The spectrum at $\ell = 1.797$ in Fig.~\ref{fig:freq_rad_spectrum_l_1.797} features strong convective signal at low frequencies, particularly close to the convective boundary. Again, the low frequency modes have the highest velocities. This only translates to high velocities at the top of the simulation domain for the modes that are most p-dominated, corresponding to those modes closest in frequency to the ``unmixed" p-mode, or $\pi$-mode \citep{Aizenman1977}. The velocities at the top of the domain for the modes in between these p-dominated modes are much smaller. The p-dominated modes with $\nu \gtrsim 200\ \mu\mathrm{Hz}$ have comparable velocities at the top to the ones of the g-dominated low frequency modes, even though the high frequency p-dominated ones are the ones seen in observations. The linearly-predicted frequencies of the g-dominated modes in the frequency range $[60, 240]\ \mu\mathrm{Hz}$ show discrepancies of a few $\mu\mathrm{Hz}$ with the simulated modes. The linearly predicted frequencies of the modes outside this frequency range and the p-dominated modes in this frequency range do not show such large discrepancies. In Fig.~\ref{fig:freq_rad_spectrum_l_4.237} with $\ell = 4.237$ the modes with largest velocities are also located at low frequencies. However, the evanescent region is sufficiently large that the g-dominated modes now barely emerge at the top of the simulation. Almost all the modes in this figure are effectively unmixed, with the mode inertia almost entirely concentrated in either the p-mode region or the g-mode region. Some modes, like the one around $\nu = 220 \ \mu\mathrm{Hz}$, do appear as mixed modes, possibly because the unmixed $\pi$- and $\gamma$-modes had similar frequencies. The frequency discrepancy between the g-dominated modes in the simulation and the linear theory prediction is larger at $\ell = 4.237$ than at $\ell = 1.797$.

\subsection{Comparison of the eigenfunctions}
\label{sec:mode_comparison}

We will now examine the agreement between the squared radial displacement eigenfunctions $|\xi_{r}|^2$ computed using linear theory and the modes found in the simulations like in e.g. \citet{Meakin2006, Alvan2014,LeSaux2025}. We examine the modes with $\ell = 0,\ 1.797,\ 2.18$ to study both the p-modes with $\ell = 0$ and the mixed modes that appear predominantly at $\ell = 1.797,\ 2.18$. The frequencies of the modes are visually identified from the spectrum. The modes are unresolved in frequency, dropping off rapidly with frequency away from the peak. Summing one bin either side of the peak proved sufficient to account for the broadening in these spectra. We directly compare the simulation radial velocity profile as a function of fractional radius with the radial displacement eigenfunctions because the radial velocity and radial displacement eigenfunctions for a given mode are related through $v_r^2 = a^2\nu^2|\xi_{r}|^2/2$ \citep{Samadi2011}, with $a$ the amplitude of that mode and $\nu$ its frequency. The eigenfunctions in velocity and displacement are therefore identical. The simulation radial velocity profiles are compared with the modes predicted by linear theory. For this comparison we want to use the linear theory mode that best matches the simulated mode.
We select this mode based on frequency proximity and matching of the mode eigenfunctions; our matching procedure is detailed in Appendix~\ref{app:mode_matching}. Extrapolation to surface velocities is detailed in Appendix~\ref{app:Extrapolation_method}. The regions where the convective signal dominates over the mode are omitted in these computations.

In Fig.~\ref{fig:modes_main} we compare the radial profile of four modes selected from the simulation, to their counterpart from linear theory, as well as their best-matching extrapolated mode. The simulation signal includes both the signal from the mode and the convective signal.

Fig.~\ref{fig:modes_pure_p_l_0.0} focuses on a pure p-mode at $\ell = 0$, $\nu = 195.5\ \mu\mathrm{Hz}$. The agreement is found to be excellent between the simulated mode in \textsc{music} and the linear theory (in fact, the curves overlap, completely obscuring the \textsc{music} one). The extrapolated mode has a slightly different frequency, and as a result the location of the nodes of the full star mode are slightly offset from the mode computed over the simulation domain. We find that, even though the amplitude of the mode itself is relatively low, the velocity at the surface is much larger due to the sharp increase of the eigenfunction towards the surface.

Fig.~\ref{fig:modes_mixed_low_freq_l_1.797} shows a mixed mode at $\ell = 1.797$, $\nu = 36.0\ \mu\mathrm{Hz}$, corresponding to a mixed f/g mode, like the ones identified in \citep{LeSaux2025}. There is a section between $r/r_\star = 0.25$ and $r/r_\star = 0.6$ where the convective signal is clearly visible in the simulated \textsc{music} signal, obscuring the mode. Everywhere else we find visual agreement between the simulated mode and linear theory. From extrapolation we see that this f/g mode has low surface velocity, even though the velocity throughout the interior is the largest of the four modes shown in Fig.~\ref{fig:modes_main}.

In Fig.~\ref{fig:modes_mixed_high_freq_l_2.18}, we show a p-dominated mode at $\ell = 2.18$, $\nu = 460.7\ \mu\mathrm{Hz}$, which again shows good agreement between theory and simulation. The extrapolated mode also shows starkly increasing amplitude towards the surface. These modes are therefore more visible than the g-dominated ones, even though in the bulk of the star the velocity of the mode is relatively low.

Finally, in Fig.~\ref{fig:modes_mixed_bad_fit_l_1.797} we show a mixed mode at $\ell = 1.797$, $\nu = 113.8\ \mu\mathrm{Hz}$. This mode falls within the frequency range $[60, 240]\ \mu\mathrm{Hz}$, where there is some discrepancy in frequency for the g-dominated modes. This particular mode is however a p-dominated mode, with a similarly small frequency discrepancy of $0.2\ \mu\mathrm{Hz}$ as the other p-dominated modes. The eigenfunctions calculated using both linear methods agree with each other in Appendix~\ref{app:Ded_GYRE_comparison}, but the mode found in the simulation disagrees with this eigenfunction. The overall structure is close to correct. The nodes are slightly offset between the simulation and linear prediction. In addition, the amplitude of the eigenfunction is too large in the radiative region, yet too small in the convective region. This discrepancy in eigenfunction also appears for the other modes in this frequency interval, both the p- and g-dominated modes. The discrepancy in both frequency and eigenfunctions are resolved when computing the modes using \textsc{GYRE} on the same radial gridpoints as used in the simulation, together with setting the numerical scheme to TRAPZ\footnote{The trapezoidal difference scheme, which is less accurate than the high-order GL6 scheme and does not evaluate the model structure at points between the grid points.}. This points to the discrepancy possibly being caused by insufficiently resolving the g-mode part of the low and intermediate frequency mixed modes in the simulation compared to the  accurate linear theory computation using GL6 and higher resolution than MUSIC. The results in the rest of this work are unchanged when using these less accurate, better agreeing modes obtained with the lower resolution, and as such we opt to maintain the accurate linear theory computation as described in Appendix~\ref{app:Dedalus_code}.

\section{Kinetic energies and surface velocities}
\label{sec:surface_amps}

 \begin{figure*}
 \centering
\subfloat[\footnotesize{The kinetic energies of the modes in the \textit{RGext90} simulation and those predicted by the empirical scaling.} \label{fig:Ekin_RGext90}]{
         \includegraphics[width=0.45\textwidth,trim={0.1cm 0.1cm 0.3cm 1.2cm},clip=true]{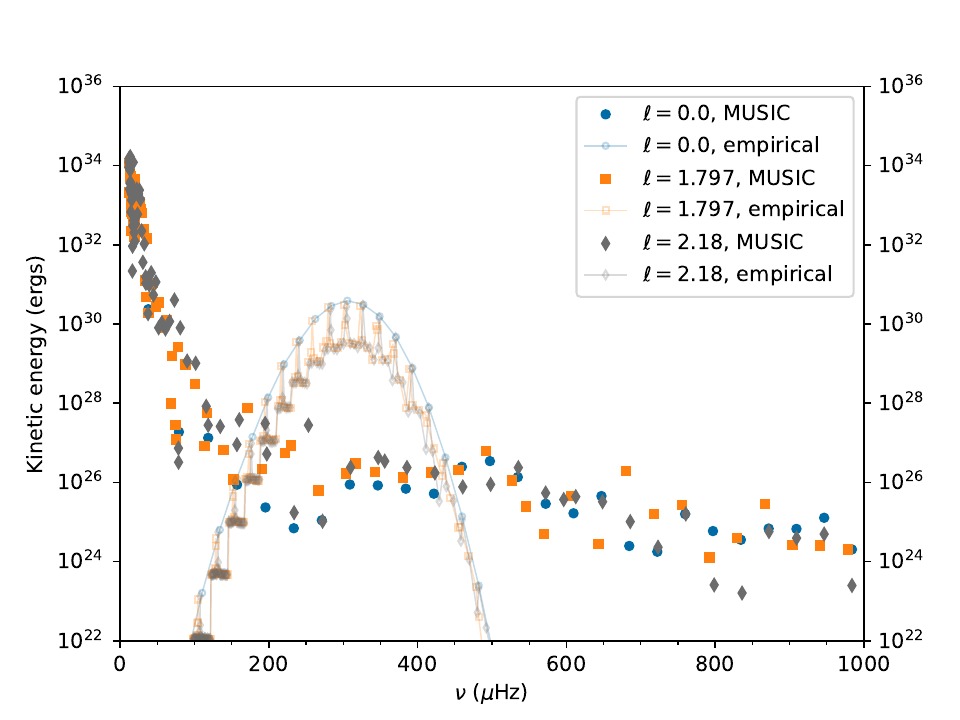}}
\hspace{0.06\textwidth}
\subfloat[\footnotesize{The kinetic energies of the modes in the \textit{RGext98} simulation and those predicted by the empirical scaling.} \label{fig:Ekin_RGext98}]{
         \includegraphics[width=0.45\textwidth,trim={0.1cm 0.1cm 0.3cm 1.2cm},clip=true]{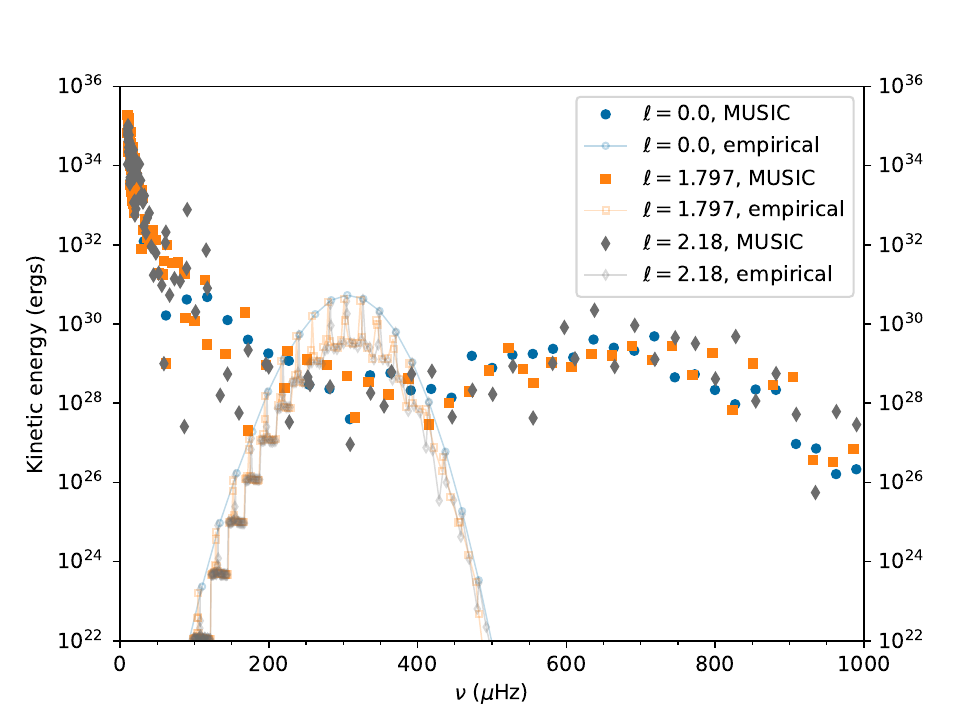}}\\
\centering
\subfloat[\footnotesize{The extrapolated surface velocities of the \textit{RGext90}  simulation and those predicted by the empirical scaling.} \label{fig:V_MUSIC_RGext90}]{
         \includegraphics[width=0.45\textwidth,trim={0.1cm 0.1cm 0.3cm 1.2cm},clip=true]{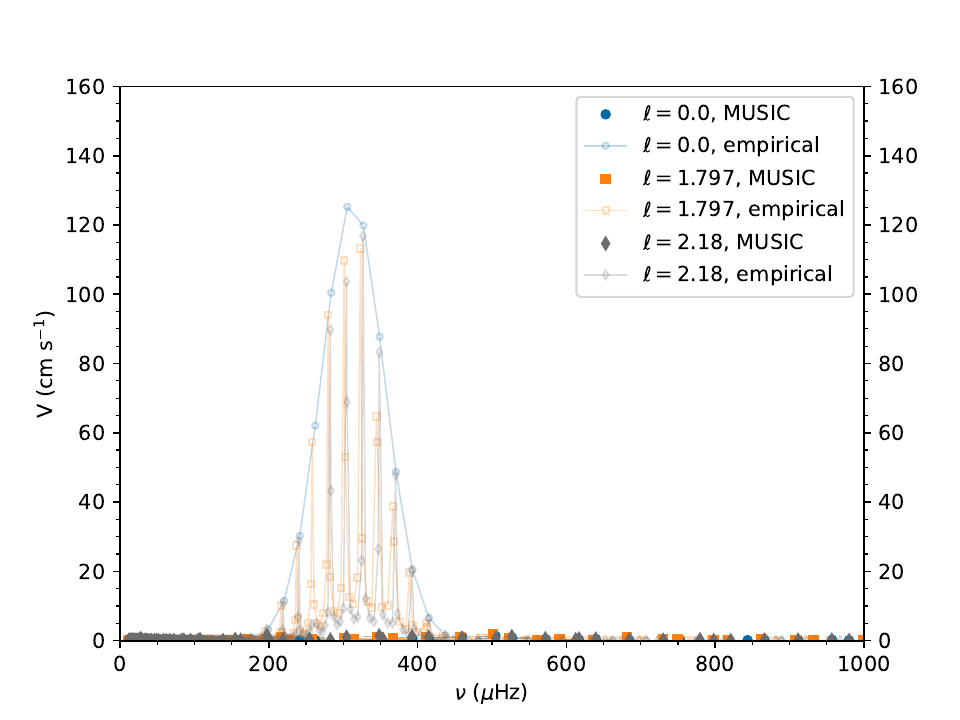}}
\hspace{0.06\textwidth}
\subfloat[\footnotesize{The extrapolated surface velocities of the \textit{RGext98}  simulation and those predicted by the empirical scaling.} \label{fig:V_MUSIC_RGext98}]{\includegraphics[width=0.45\textwidth,trim={0.1cm 0.1cm 0.3cm 1.2cm},clip=true]{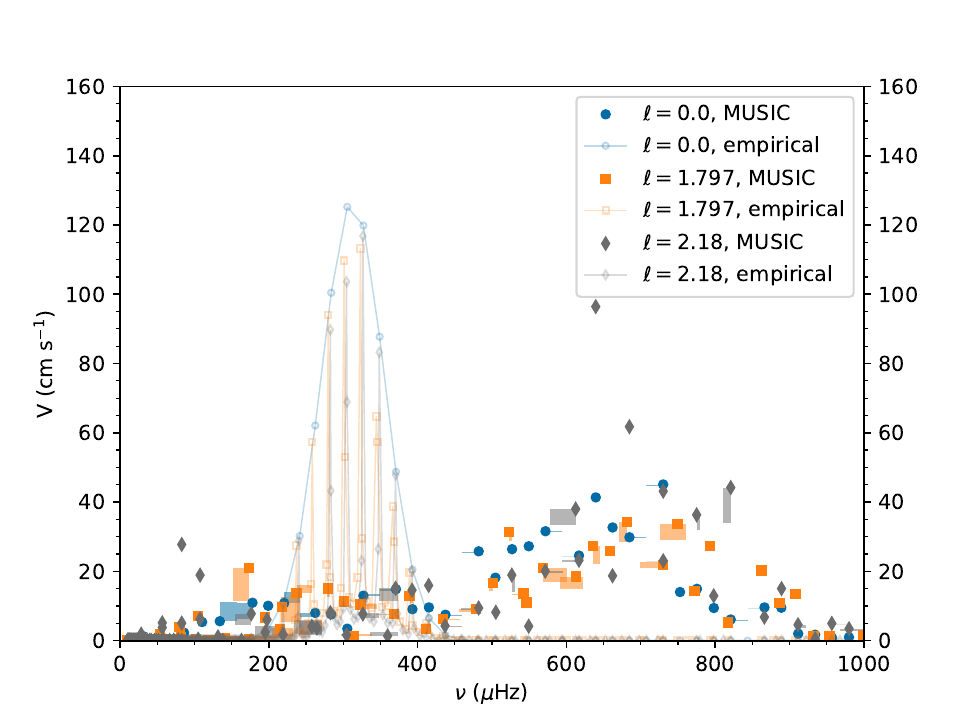}}
 \caption{Kinetic energies (top row) and extrapolated surface velocities (bottom row) of both the \textit{RGext90} (left column) and \textit{RGext98} (right column) simulations in blue circles, orange squares and grey diamonds for $\ell = 0,\ 1.797,\ 2.18$ respectively. The kinetic energies are computed within the truncated region of the simulation. Both the kinetic energies and surface velocities are compared to empirical estimates using the method of \citep{Belkacem2015b,Bordadagua2025}. These empirical estimates are plotted in smaller transparent markers joined by solid-lines. The surface velocities are extrapolated using the \textsc{gyre}  modes computed using the full stellar model, the modes selected using the method in Appendix~\ref{app:mode_matching} are plotted in solid markers. The extrapolated surface velocities of the modes with $\mathcal{Q} > 0.5$ are shown in a transparent rectangle around the marker. These rectangles give an estimate of what the surface velocities could be for the other extrapolated modes for one simulated mode.}
    \label{fig:amp_plots}

\end{figure*}

In this section we reintroduce the \textit{RGext98} simulation to study the effect of radial truncation on the modes. We study the volume-integrated kinetic energy in the simulated region and the extrapolated surface velocities as a function of frequency for $\ell < 3$. The volume-integrated kinetic energy for a mode with frequency $\nu$ and harmonic degree $\ell$ is defined as:

\begin{equation}
    E_{\nu,\ell} \equiv 2\pi\int_{r_i}^{r_o} \rho\left(P[\hat{v}_r](\nu,\ell)/2 + P[\hat{v}_\theta](\nu,\ell)/2\right) r^2 \: \mathrm{d}r,
\end{equation}
where we note that $P[\hat{v}_r]/2 = v_r^2$ and $P[\hat{v}_\theta]/2 = v_\theta^2$. This kinetic energy corresponds to that of a full 3D spherical shell of the truncated region. We compare the simulation results to those found using the empirical scaling of \citet{Belkacem2015b,Bordadagua2025}, which mimic those of observations. More details on this empirical scaling method can be found in Appendix~\ref{app:Belkacem_explanation}. 
A stark difference between the simulation and empirical kinetic energies is immediately apparent in Fig.~\ref{fig:Ekin_RGext90}. Instead of the empirical Gaussian shape---chosen to represent observations---centred around the frequency of maximum amplitude $\nu_\mathrm{max}' = 312.8\ \mu \mathrm{Hz}$, the kinetic energies of the simulated modes are largest at low frequencies. They then drop off rapidly as a function of frequency. At frequencies $\nu < 50\ \mu\mathrm{Hz}$ the kinetic energies in the simulations are larger than those predicted by the empirical scaling. Examining Fig.~\ref{fig:Ekin_RGext98} we find that the simulated kinetic energies of \textit{RGext98} are larger than those of \textit{RGext90}. The increased convective velocities found throughout the simulation domain in \textit{RGext98} have in turn increased the velocities of the modes and thus their kinetic energies. In Fig.~\ref{fig:Ekin_RGext98} a wider Gaussian-like shape of modes, centred around $\nu = 700\ \mu\mathrm{Hz}$, has emerged for \textit{RGext98}. Excitation of p-modes takes place close to the surface \citep[e.g.][]{Goldreich1977,Goldreich1990,Samadi2003}, and the mixed modes are also expected to be excited there \citep{Samadi2011}, which likely explains the emergence of the Gaussian-like shape only in the simulation with outer boundary close to the surface. The behaviour of the highly-energetic low frequency modes, and in particular the sharp decrease with increasing frequency, is similar for both simulations. These modes might therefore be excited at the convective boundary, the increased velocity of the modes explained by the larger velocities there in the \textit{RGext98} simulation.\\

Next we turn to the surface velocities $V = \sqrt{v_r^2 + v_\theta^2}$ that we extrapolate from the simulation results using the \textsc{gyre} modes. We compute the surface velocities for each mode in the simulation using the matching and extrapolation method already employed in Fig.~\ref{fig:modes_main}, detailed in Appendices~\ref{app:mode_matching} and \ref{app:Extrapolation_method}.
For each simulated mode, the best matching mode from the full stellar model is plotted at its own frequency and extrapolated surface velocity (filled markers). Note that our extrapolation ensures that the total kinetic energy in all plotted modes matches the total kinetic energy in the simulation.

Around each simulation mode, there are multiple candidate modes of the full stellar model with similar frequency and eigenfunction.
The mode matching process can therefore impact the determination of surface velocities.
For each simulation mode, we thus identify ``well-matching modes'':
those are the candidate full stellar model modes of Appendix~\ref{app:mode_matching} that additionally verify $\mathcal{Q} > 0.5$
(i.e. whose eigenfunction matches better than 50\%). In Fig.~\ref{fig:amp_plots}, along with the marker of each best-matching mode,
we also plot the bounding box of frequencies and velocities of its ``well-matching modes''. This helps visualize the impact of mode matching on the inferred surface velocities.

In addition to these extrapolated simulation results, the empirical surface velocities are plotted in Fig.~\ref{fig:V_MUSIC_RGext90} and Fig.~\ref{fig:V_MUSIC_RGext98}, to illustrate the expected surface velocities in observations.

The extrapolated surface velocities for \textit{RGext90} in Fig.~\ref{fig:V_MUSIC_RGext90} are much smaller than expected from the empirical scaling law, in accordance with the smaller kinetic energies compared to the empirical results. Most extrapolated surface velocities are smaller than $1\ \mathrm{cm}\ \mathrm{s}^{-1}$, with a peak value of $\approx 2 \mathrm{cm}\ \mathrm{s}^{-1}$. We thus find no significant surface presence of the modes based on extrapolation of this model. In Fig.~\ref{fig:V_MUSIC_RGext98} significant surface velocities are present for the high frequency modes between $500\ \mu\mathrm{Hz}$ and $800\ \mu\mathrm{Hz}$, in a somewhat Gaussian-like shape. The solution spaces offer small variations in surface velocity, which are not large enough to alter this pattern. Between $200\ \mu\mathrm{Hz}$ and $400\ \mu\mathrm{Hz}$, where the empirical modes lie for this stellar model, we also find non-negligible surface velocities, but not as large or in the Gaussian shape predicted by the empirical scaling. The modes with highest surface velocities are thus shifted to higher frequencies compared to the empirical scaling. This shift could be caused by a number of factors not yet included in our simulations. For example, inclusion of the uppermost and lowermost parts of the star, 3D convection, rotation and magnetic fields, or different boundary conditions, could all affect the excitation and damping of these modes, or even their presence in the simulation. Finally, the surface velocities of most low frequency ($<100 \mu\mathrm{Hz}$) modes are very small. A few modes, those which are most p-dominated, have amplitudes in the range $5$ to $10$ $\mathrm{cm}\ \mathrm{s}^{-1}$, with one reaching $30\ \mathrm{cm}\ \mathrm{s}^{-1}$. These would be easiest to observe, allowing confirmation of the amplitude of these low frequency modes in the stars themselves, but their detections are further complicated by the strong convective signal present at these low frequencies.

\section{Conclusions and discussion}
\label{sec:conclusions}

In this work we present, for the first time, 2D hydrodynamical simulations of mixed modes in a red giant star using \textsc{music}. We have performed two simulations of the M1 red giant star \citep{Belkacem2015b} in a meridional wedge, aiming to investigate the amplitudes of the mixed modes, in relation to their efficiency in transporting angular momentum (AM). These simulations, \textit{RGext90} and \textit{RGext98}, are truncated at fractional inner radius $r_i/r_\star = 0.02$ and fractional outer radii $r_o/r_\star=0.90$ and $r_o/r_\star=0.98$ respectively. We have compared the frequencies and eigenfunctions of all modes identified in our simulations with those predicted by \textsc{gyre}  as well as our own eigenvalue solver written using \textsc{dedalus}. We find that the frequencies match very well for p-dominated modes, while the g-dominated modes, especially those in the frequency range $[60, 240]\ \mu\mathrm{Hz}$, tend to be mismatched by one to a few $\mu\mathrm{Hz}$. Equally, we find excellent agreement between the simulated radial velocity profiles and the linear theory radial displacement eigenfunctions except for the modes within the aforementioned frequency range, for both p- and g-dominated modes. We find that the kinetic energies of the modes around $\nu_\mathrm{max}' = 312.8\ \mu \mathrm{Hz}$ in the \textit{RGext90} simulation are several orders of magnitude smaller than those predicted by the empirical scaling relations of \citet{Belkacem2015b}. The modes at frequencies $\nu <50\ \mu\mathrm{Hz}$ have kinetic energies much larger than obtained from the empirical prediction, which rapidly decrease with increasing frequency. The kinetic energies as a whole are larger in the \textit{RGext98} simulation. In addition, a broad Gaussian-like shape of large kinetic energies has emerged around $\nu = 700\ \mu\mathrm{Hz}$. Extrapolating the modes in both simulations shows that the highly-energetic modes with $\nu <50\ \mu\mathrm{Hz}$ have low surface velocities, making them hard to detect. The modes around $\nu = 700\ \mu\mathrm{Hz}$ show the observationally expected Gaussian-like shape only for the \textit{RGext98} simulation, while they have negligible surface velocities for the \textit{RGext90} simulation. The extrapolated surface velocities of the high frequency \textit{RGext98} modes, situated at frequencies larger than predicted by the empirical scaling, are comparable to those found from the empirical scaling relations. The AM transported by these previously unknown highly-energetic low frequency modes has not yet been investigated, so far only the higher frequency empirical Gaussian has been taken into account in the prescription of \citet{Belkacem2015b}. These low frequency modes might be sufficient to explain the small amount of AM transport required in young red giants \citep{Eggenberger2017} and the larger amount of AM transport for very evolved red giants as the density of modes and their amplitudes increases, particularly at such low frequencies. Likewise, in subgiants the AM transport by mixed modes has been ruled out as an explanation for the slow development of the differential rotation \citep{Belkacem2015b}. In these stars the evanescent region at low frequencies is small, such that these low frequency highly-energetic modes may again be relevant to transport AM, even though the density of mixed modes is lower than in red giant stars.

Future simulations should analyse different evolutionary stages, from subgiants to more evolved red giants, to further understand the effect of stellar evolution on mixed modes and their amplitudes. To improve the extrapolation of modes in the simulation, allowing better constraints on observable quantities such as the surface velocity as well as constraining the AM transport by mixed modes by better understanding their amplitudes, we require a better understanding of the excitation of waves and therefore standing modes in the simulations. One possibility would be to attempt to verify the excitation schemes detailed in e.g.~\citet[]{Samadi2001, Belkacem2008, Bessila2024} for excitation in the convection zone and those detailed in e.g.~\citet[][]{Lecoanet2013, Pincon2016} for the excitation at the convective boundary. This would also require constraining the damping rate of the modes in simulations. A study of the meridional extent of the wedge slice, which in turn changes the size of the evanescent region in these simulations by changing the possible values of $\ell$, should also be performed to understand the effects of the size of the evanescent region on the mixed mode amplitudes, as well as assist in extrapolation from wedge simulations to full stars. Finally, the present results include a number of approximations (2D axisymmetry, wedge slices, radial truncation) which should be relaxed, as well as missing physics (no rotation, magnetic field) which should be included in future work. In particular, the 2D axisymmetry and absence of rotation means that AM transport cannot be directly studied in these simulations. Performing these same simulations in 3D would allow to study the AM transport present, and would possibly allow to connect the observed AM transport to the mixed modes amplitudes. In such 3D simulations the flow velocities are expected to be smaller than in 2D, which therefore is likely to reduce the amplitudes of the mixed modes observed in this work. In addition, 3D convection is also smaller scale than 2D convection, which therefore might change the scales at which waves are excited. Even with these approximations, this work clearly demonstrates the feasibility of compressible hydrodynamical simulations to capture mixed modes in stars, unlocking opportunities to determine the amplitude of these modes and provide reliable, quantitative estimates of the efficiency of angular momentum transport as well as potentially predicting the observability of mixed modes ahead of the upcoming PLATO mission.

\section*{Data availability}

The kinetic energies and surface velocities shown in Fig.~\ref{fig:amp_plots} as well as the underlying spectral data of this work can be found in a Zenodo repository at\dataset[10.5281/zenodo.18661976]{10.5281/zenodo.18661976}.

\begin{acknowledgments}
We would like to thank the referee for their careful reading of the manuscript and their constructive comments that helped improve the paper. NBV would like to thank K. Belkacem and J. Philidet for helpful discussions. NBV is supported by STFC grant ST/Y002164/1. ALS acknowledges support from the European Research Council (ERC) under the Horizon Europe programme (Synergy Grant agreement 101071505: 4D-STAR) from the CNES SOHO-GOLF and PLATO grants at CEA-DAp, and from ATPS (CNRS/INSU). Part of this work was supported by the ERC grant No. 787361-COBOM. RHDT acknowledges support from NASA grants 80NSSC24K0895 and 80NSSC23K1517, and NSF grant 2407636. AL is supported by ERC Starting Grant 101165631 (``Calcifer"). The authors would like to acknowledge the use of the University of Exeter High-Performance Computing (HPC) facility ISCA in carrying out this work. This work used the DiRAC Memory Intensive service (Cosma8) at Durham University, managed by the Institute for Computational Cosmology, and the DiRAC Data Intensive service (DIaL3) at the University of Leicester managed by the University of Leicester Research Computing Service. These facilities are managed on behalf of the STFC DiRAC HPC (www.dirac.ac.uk). The DiRAC services at Durham and Leicester were funded by BEIS, UKRI and STFC capital funding, and STFC operations grants. The service at Durham received funding from Durham University. DiRAC is part of the UKRI Digital Research Infrastructure. 
\end{acknowledgments}

\begin{contribution}

\
NBV was responsible for running the simulations, developing the analysis software, analysing the results, writing the original draft of the manuscript and editing the final draft. 
ALS came up with the initial research concept, performed the proof of concept investigation, contributed to results analysis and edited the manuscript.
IB came up with the initial research concept, contributed to results analysis, obtained the funding and edited the manuscript.
TG developed the MUSIC software, contributed to results analysis, obtained the funding and edited the manuscript.
RHDT developed a modified version of the GYRE software, helped with linear analysis and edited the manuscript.
AL developed the Dedalus eigenvalue solver software, helped with linear analysis and edited the manuscript.
AM developed the MUSIC software and edited the manuscript.

\end{contribution}

%



\appendix

\section{Numerical setup of the simulations in \textsc{music}}

\subsection{Code and equations}
\label{app:MUSIC_code}
The simulations in this work have been performed using the time-implicit code \textsc{music}, which solves the continuity, momentum and internal energy equations:

\begin{align}
    \frac{\partial \rho}{\partial t} &= - \nabla \cdot (\rho \bm{v}),\\
    \frac{\partial \rho \bm{v}}{\partial t} &= -\nabla\cdot (\rho \bm{v} \otimes\bm{v}) - \nabla p + \rho \bm{g},\\
    \frac{\partial \rho e}{\partial t} &= -\nabla \cdot (\rho e \bm{v}) - p\nabla \cdot \bm{v} + \nabla \cdot (\chi \nabla T) + \rho \varepsilon_{\mathrm{nuc}},
\end{align}
with $\rho$ the density of the fluid, $e$ the specific internal energy and $\bm{v}$ the velocity. The gravity is given by $\bm{g} = g(r)\hat{\bm{r}}$ and solved self-consistently in 1D from $\langle\rho\rangle_\theta$ within each implicit time step. The temperature $T$ and pressure $p$ are obtained from $\rho$ and $e$ using the tabulated equations of state found in the OPAL tables \citep{Rogers2002}. The radiative diffusion is computed using the heat conductivity $\chi$, calculated from the Rosseland mean opacity found in the OPAL opacity tables \citep{Iglesias1996}. Both the equation of state and opacity tables are retrieved from MESA version r15140. The heating from nuclear burning $\varepsilon_{\mathrm{nuc}}$ is taken from the 1D initial model, and is treated as constant during the 2D simulation as the timescale of evolution of the nuclear burning is much longer than the timescale of the run.

\subsection{Domain and boundary conditions}
Choosing a wedge simulation domain allows using periodic instead of reflective boundary conditions on the horizontal boundaries.
This avoids hard reflective walls, letting waves and standing modes propagate more freely,
while also avoiding horizontal boundary layers in the simulation.
In the radial direction we use reflective boundary conditions for the velocity, which correspond to $v_r = 0$ and $\frac{\partial}{\partial r} v_\theta = 0$ at $r_{i}$ and $r_o$. The energy fluxes at both boundaries are taken from the energy fluxes at corresponding radii in the 1D initial model. The density is linearly extrapolated into the ghost cells at the inner boundary and logarithmically extrapolated at the outer boundary. The simulation grid has uniform spacing in the $r$ and $\theta$ directions. The radial spacing is chosen such that there are at least 140 radial grid cells per pressure scale height at the convective boundary, as justified in \cite{Baraffe2023}.

\subsection{Spectral analysis}
For spectral analysis, the functions are expanded into wedge harmonics, i.e. eigenfunctions of the Laplacian on our wedge domain.
Their associated wavelengths and eigenvalues differ from the spherical harmonics (which are eigenfunctions of the Laplacian on the full sphere).
We therefore label the wedge harmonics by their ``effective'' degree $\ell$ such that their eigenvalue is $\ell (\ell + 1)$,
to relate them to spherical harmonics which probe a similar angular scale.
More details on the wedge harmonics can be found in \cite{LeSaux2025}.
For the wedge with our chosen $\theta_i$ and $\theta_o$, the lowest non-zero effective harmonic degree is $\ell = 1.797$.

\subsection{Time integration}
We use the second-order time integrator TR-BDF2 \citep{Hosea1996} after reaching steady state, to reduce the damping effect of the time integrator. The maximum hydrodynamic Courant number (relative to the sound speed) in steady state is set to 100, and the maximum advective Courant number (relative to the flow velocity) to 0.1.

\subsection{Initial stellar models}

 \begin{figure}
     \centering
     \includegraphics[width=0.45\textwidth]{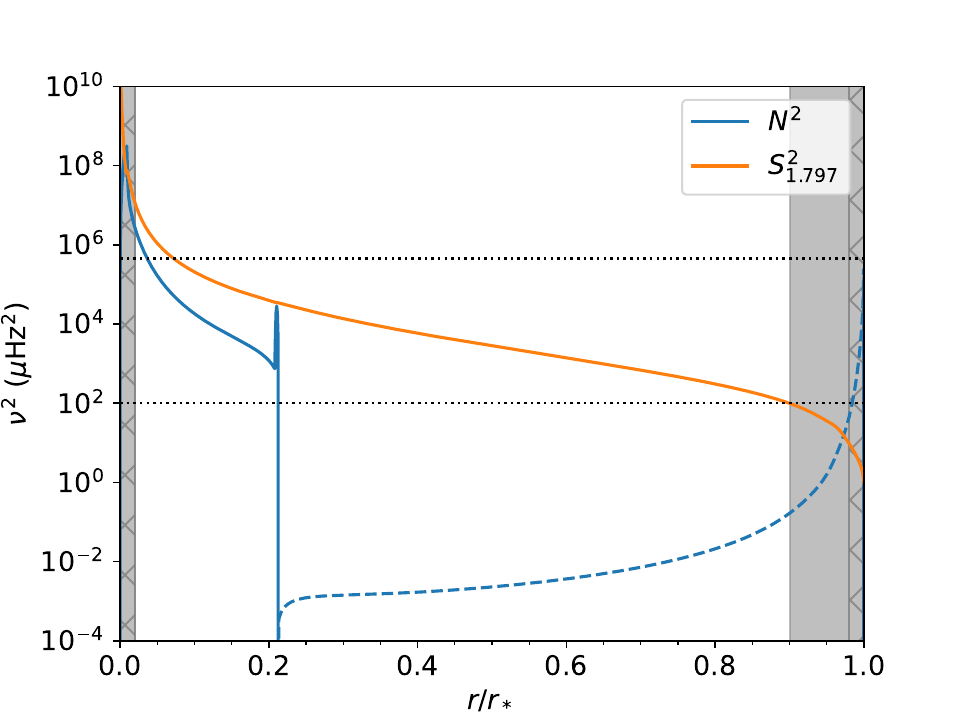}
     \caption{The 1D input model for our simulations of the M1 red giant star \citep{Belkacem2015b} generated using MESA. The squared buoyancy frequency is given in blue, solid- and dashed-blue represent positive and negative squared buoyancy frequency respectively. The squared Lamb frequency for $\ell =1.797$, corresponding to the lowest non-zero $\ell$ in the wedge simulations, is given in orange. In the model \textit{RGext90} the region between the light grey shadings is simulated, while in the model \textit{RGext98} the region between the cross-hatched dark greys is simulated. Modes with frequencies between the horizontal dotted lines are examined, with the upper dotted line corresponding to the Nyquist frequency.}
    \label{fig:initial_model}
 \end{figure}

The initial stellar models are computed using version r24.08.1 of the stellar evolution code \textsc{mesa} \citep{Paxton2011,Paxton2013,Paxton2015,Paxton2018,Paxton2019,Jermyn2022}. The initial helium abundance is set to $Y = 0.28$ and the metallicity to $Z=0.02$. These initial models are computed using the Schwarzschild criterion for the onset of convection, without overshooting, mass-loss or extra mixing. The squared buoyancy frequency $N^2$ and the squared Lamb frequency $S^2_{\ell} = \ell (\ell + 1) c_s^2/r^2$, with $\ell$ the harmonic degree, of the chosen initial model as a function of radius are shown in Fig.~\ref{fig:initial_model}. The MUSIC initial conditions are obtained from this 1D initial model, where we have ensured that it reproduces the $N^2$ profile of the 1D MESA model whilst maintaining hydrostatic equilibrium. This results in minor deviations in the temperature and density profiles. For both simulations the fractional deviations from the 1D MESA model are $\mathcal{O}(10^{-3})$ in most of the domain, peaking towards the inner boundary with a fractional deviation of $\approx0.015$.
We study the mixed modes in the frequency range $\nu = \omega/(2\pi) \in [10, 666.67]\ \mu \mathrm{Hz}$, demarcated by the two horizontal black-dotted lines. This frequency range covers the entirety of the buoyancy glitch, which results from the first dredge-up, and thus ensures that the modes which interact with the buoyancy glitch are included. The expected frequency of the mode with maximum amplitude, $\nu_\mathrm{max}' = 312.8\ \mu \mathrm{Hz}$, found using scaling relations described in Appendix~\ref{app:Belkacem_explanation}, lies in the middle of the studied region. The prime indicates that it is a quantity of the full, non-truncated, stellar model. We also include in this study the spectrally reflected modes between the Nyquist frequency of $666.67\ \mu\mathrm{Hz}$ and $1000\ \mu\mathrm{Hz}$.  The large frequency separation of this stellar model is $\Delta\nu' = 22.79\ \mu\mathrm{Hz}$ and the period spacing for $\ell = 1.797$ is $\Delta \Pi_{1.797}' = 58.86\ \mathrm{s}$.

\section{Linear theory} 
\label{app:linear_theory}

\subsection{Setup of the \textsc{dedalus} eigenvalue solver}
\label{app:Dedalus_code}
The eigenvalue solver constructed using the \textsc{eigentools} package \citep{Eigentools} for \textsc{dedalus} v2 \citep{Burns2020} is used to calculate the frequencies and radial mode profiles of the mixed modes. We follow the method in \cite{Perrot2019, Leclerc2022, Leclerc2024}. First the compressible linearised Euler equations for the perturbed quantities $v_r,\ v_\theta,\ \rho,\ p$ and $\Phi$, the gravitational potential, are rewritten into an eigenvalue problem with a linear operator. We solve the linearised equations for mass conservation:

\begin{equation}
    \frac{\partial \rho}{\partial t} = -\bm{v}\cdot\nabla\rho_0 - \rho_0\nabla\cdot\bm{v},
\end{equation}
momentum conservation:

\begin{equation}
    \rho_0\frac{\partial \bm{v}}{\partial t} = -\nabla p + \rho \bm{g}_0 + \alpha_g\rho_0\bm{g},
\end{equation}

and an equation of state, $p=p(\rho,s)$, with $s$ the entropy. Adiabaticity is assumed, i.e. $\mathrm{d}s=0$, which yields:
\begin{equation}
    \frac{\partial \rho}{\partial t} + \bm{v}\cdot\nabla\rho_0 = \frac{1}{c_s^2}\left(\frac{\partial p}{\partial t} + \bm{v}\cdot\nabla p_0\right),
\end{equation}
with $\rho_0$, $p_0$ and $\bm{g}_0$ the unperturbed density, pressure and gravity and $\alpha_g$ a parameter set to either zero or one to solve the equations with or without the Cowling approximation \citep{Cowling1941}. Forgoing this approximation requires additionally solving for the gravitational potential using Poisson's equation:

\begin{equation}
    \nabla^2\Phi = 4\pi G \rho,
\end{equation}
with $G$ the gravitational constant. It is important to stress that these equations are fully adiabatic and no diffusion is included. The equations are linearised and the same coordinate substitution as in \cite{Leclerc2022} is performed:

\begin{align}
    \tilde{v}_r = \rho_0^{1/2}r v_r,\\
    \tilde{v}_\theta = \rho_0^{1/2}r v_\theta,\\
    \tilde{p} = \rho_0^{-1/2}c_s^{-1}rp,\\
    \tilde{\Theta} = \rho_0^{-1/2}r\frac{g}{N}\left(\rho - \frac{1}{c_s^2}p\right)
\end{align}
No coordinate substitution is performed for $\Phi$. Next we project to vector spherical harmonics as in \citet{LeSaux2025} and solve the following eigenvalue problem:

\begin{equation}
    \nu A \bm{X} = \mathcal{H}\bm{X},
\end{equation}
with
\begin{widetext}
\begin{equation}
    \mathcal{H} =
    \begin{pmatrix}
        0 & 0 & 0 & S_{\ell} & \alpha_\mathrm{g}\rho_0^{1/2}rS_\ell/c_s\\
        0 & 0 & iN & i\left(-S + c_s\partial_r + \frac{1}2{}\frac{\mathrm{d}c_s}{\mathrm{d}r}\right) & \alpha_\mathrm{g}i\rho_0^{1/2}r \partial_r\\
        0 & -iN & 0 & 0 & 0\\
        S_\ell & i \left(S + c_s \partial_r + \frac{1}{2}\frac{\mathrm{d}c_s}{\mathrm{d}r}\right) & 0 & 0 & 0\\
        0 & 0 & \alpha_\mathrm{g}4\pi G\rho_0^{1/2}/r (N/g_0) & \alpha_\mathrm{g}4\pi G\rho_0^{1/2}/(rc_s) & \alpha_\mathrm{g}\left(S_\ell^2/c_s^2 - 1/r^2\partial_r(r^2\partial_r)\right)
    \end{pmatrix},
\end{equation}
\end{widetext}
and $\bm{X} = (\tilde{v}_r, \tilde{v}_\theta, \tilde{\Theta},\tilde{p},\Phi)^T$, $A = \mathrm{diag}(1,1,1,1,0)$, $\nu$ the (complex) frequency of the mode, $S = \frac{c_s}{2g}\left(N^2-\frac{g^2}{c_s^2}\right) - \frac{1}{2}\frac{\mathrm{d}c_s}{\mathrm{d}r}+\frac{c_s}{r}$ the buoyancy-acoustic frequency. We use the same non-integer values of $\ell$ that result from the wedge harmonic transform of the simulation data. Furthermore, we set the same boundary conditions on the radial velocity that we specify in the 2D simulations within this eigenvalue solver, i.e. $v_r(r_i) = v_r(r_o) =0$. Furthermore, if the Cowling approximation is not employed we add the regularity condition at the inner boundary 
\begin{equation}
    \ell\Phi(r_i) - r_i\frac{\mathrm{d}\Phi}{\mathrm{d}r}\Bigr|_{r_i} = 0,
\end{equation}
and vanishing potential on the outer boundary 
\begin{equation}
    (\ell+1)\Phi(r_o) + r_o\frac{\mathrm{d}\Phi}{\mathrm{d}r}\Bigr|_{r_o}  = 0.
\end{equation}

We solve these equations using a spectral method, decomposing the solution using the Chebyshev polynomial basis with a total number of grid points $N_R$. We use the \textsc{eigentools} drift threshold functionality which retains the eigenmode solutions only if the difference in frequencies between the problem solved on the grid with $N_R = 1024$ and $N_R = 1536$ divided by the lower resolution frequency is less than the chosen value of $10^{-2.5}$. For more detail on the drift threshold method see e.g. \citet[][]{boyd2001chebyshev}. We store the frequency and mode profiles corresponding to the $N_R = 1536$ solution for all retained modes. We perform the exact same computations using \textsc{gyre} as with the \textsc{dedalus} eigenvalue solver. For the modes of truncated model we set both the inner and outer boundary condition to ZERO\_R to match the simulations. We compute the frequencies using the adiabatic calculation and COLLOC\_GL6, i.e. sixth-order Gauss-Legendre collocation.

\textsc{music} 7.0.0 solves Poisson's equation inside each implicit step using horizontal averages of the density resulting in one-dimensional gravity, i.e. $\bm{g} = g(r)\hat{\bm{r}}$. For $\ell > 0$ modes the horizontal average of the density perturbation is zero, which results in the perturbation to the gravitational potential going to zero. This is effectively equivalent to the Cowling approximation for $\ell > 0$ in the 2D simulation. To compare to the 2D simulations the Cowling approximation must therefore be used to compute the $\ell > 0$ modes using linear theory. The perturbation to the gravitational potential for $\ell = 0$ modes is fully radial. As a result these modes do not feel the effect of the horizontal averaging and thus need to be computed without the Cowling approximation to match the simulation modes.\\
The computations of the linear modes of the full stellar model are performed using \textsc{gyre} only. The inner boundary condition is set to REGULAR and outer boundary condition to VACUUM, i.e. vanishing surface density. These modes have been computed without the Cowling approximation regardless of the value of $\ell$, so they best correspond to the modes that are present in stars.

\subsection{Comparison of \textsc{dedalus} and \textsc{gyre}}
\label{app:Ded_GYRE_comparison}

\begin{figure}
    \centering
    \includegraphics[width=0.45\textwidth]{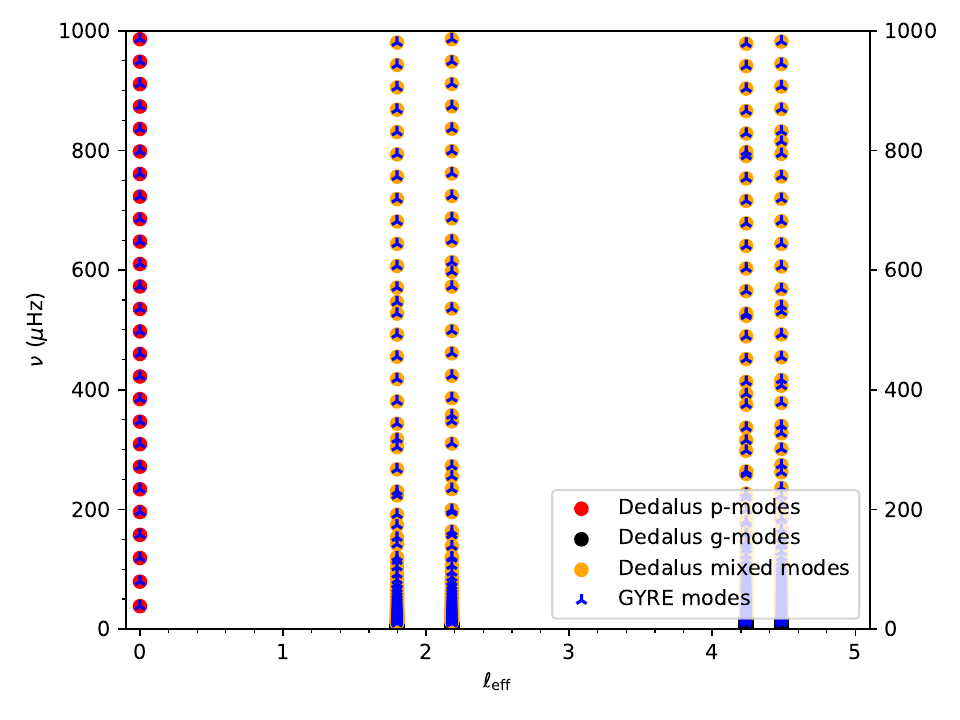}
    \caption{Comparison of the frequencies of the modes obtained using the \textsc{dedalus} eigenvalue solver plotted in red, black and orange circles, corresponding to p-, g- and mixed modes respectively. The frequencies found using \textsc{gyre} are over plotted in blue triangles. }
    \label{fig:Gyre_Ded_freq_comp}
\end{figure}

We compare the frequencies of all modes computed using both \textsc{gyre}  and \textsc{dedalus} as well as the radial displacement eigenfunctions of two example modes to examine the agreement between the two methods. The frequencies as a function of $\ell$ are compared in Fig.~\ref{fig:Gyre_Ded_freq_comp}. The agreement between \textsc{dedalus} and \textsc{gyre} for these studied values of $\ell$ is excellent.\\

\begin{figure}
\subfloat[The linear theory mode with $\ell = 0$, $\nu = 195.6\ \mu\mathrm{Hz}$ . \label{fig:linear_theory_pmode}]{\includegraphics[width=0.45\textwidth]{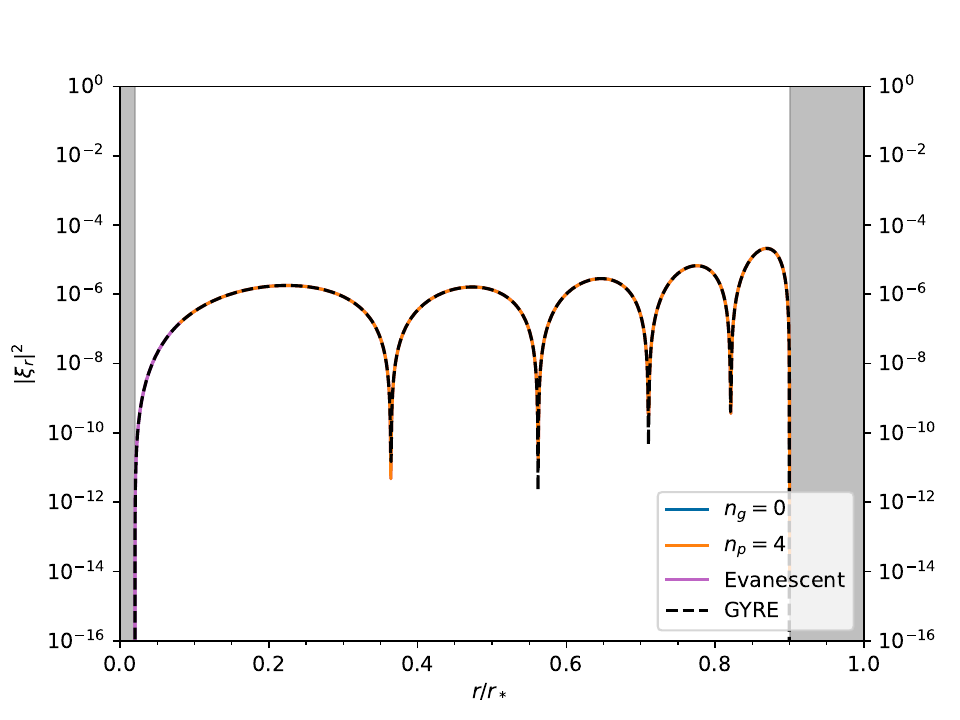}}\\

\subfloat[The linear theory mode with $\ell = 1.797$, $\nu = 114.0\ \mu\mathrm{Hz}$. \label{fig:linear_theory_mixedmode}]{\includegraphics[width=0.45\textwidth]{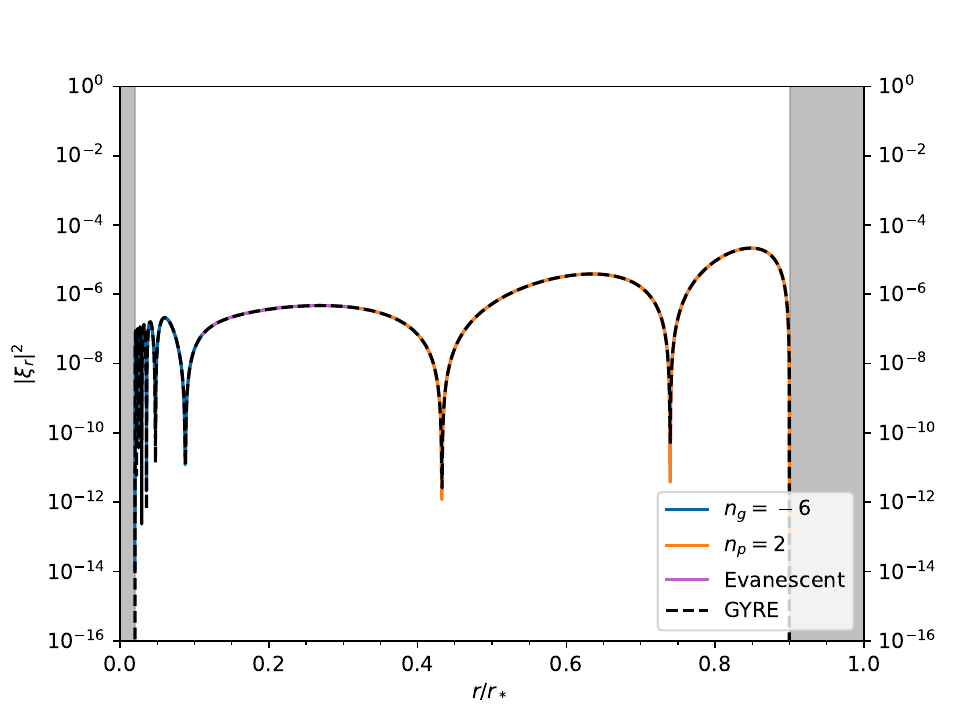}}

     \caption{Comparison of the radial displacement eigenfunctions obtained from linear theory using the \textsc{dedalus} eigenvalue solver and \textsc{gyre}. The eigenfunction obtained from the \textsc{dedalus} eigenvalue solver is plotted in solid-blue, -orange and -purple, corresponding to the g-mode, p-mode, and evanescent regions respectively. The \textsc{gyre} eigenfunction is overlaid in dashed-black, showing excellent agreement for the modes.}
    \label{fig:GYRE_Ded_modes_comp}

\end{figure}

The comparison of the radial displacement eigenfunctions is shown in Fig.~\ref{fig:GYRE_Ded_modes_comp}. The solid-blue, orange and purple lines are the predictions obtained from the \textsc{dedalus} eigenvalue solver, corresponding to the regions where the modes have g-mode, p-mode character and where they are evanescent. The \textsc{gyre}  profile is overlaid in dashed-black, showing excellent agreement for both example modes. All other modes show the same excellent agreement. Evidently, the two agree on both frequency and mode profiles for both $\ell = 0$ and $\ell >0$. Thus we can be confident that the results from linear theory obtained through either method are robust.

\subsection{Matching modes between simulation and full stellar model}
\label{app:mode_matching}

To extrapolate the simulated modes to the surface we wish to select the mode of the full stellar model that most resembles the simulated mode in the simulation domain. The $\Delta\nu'$ and $\Delta\Pi_{\ell}'$ of the full stellar model are smaller than those of the simulations, so there are more mixed modes in the full stellar model than in the simulation.

We therefore need a procedure to assign, for each mode present in the simulation (denoted without a prime), one or more extrapolated modes in the full stellar model (denoted with a prime); for this, we match the eigenfunctions of the modes. Following e.g. \cite{Townsend2025}, we introduce the mass-weighted inner product between two eigenvectors $\bm{f},\ \bm{g}$:
\begin{equation}
    \left< \bm{f}, \bm{g} \right> = \int_{r_i}^{r_o} \bm{f}^* \cdot\bm{g} \rho r^2\: \mathrm{d}r.
\end{equation} 
The integral spans the simulated region only, where both simulated and full stellar model modes are defined; this inner product can thus be computed for any combination of primed and unprimed arguments. We then define the $L^2$-norm of a mode $\bm{\xi}$ (primed or unprimed) as:
\begin{equation}
    \| \bm{\xi} \| = \sqrt{\left< \bm{\xi}, \bm{\xi} \right>}.
\end{equation}
We quantify the degree of match between a simulated mode $\bm{\xi}$ and a full stellar model mode $\bm{\xi'}$ using:
\begin{equation}
    \mathcal{Q}(\bm{\xi'};\bm{\xi}) = \frac{\left| \left< \bm{\xi'}, \bm{\xi} \right> \right|}{\|\bm{\xi'}\| \|\bm{\xi}\|}
\end{equation}
which is a real number between 0 (orthogonal modes) and 1 (perfect match). For each simulated mode $\bm{\xi}$ with frequency $\nu$, we first select all the full stellar model modes $\bm{\xi'}$ whose frequency $\nu'$ is in $[\nu - \Delta\nu', \nu + \Delta\nu']$. We then compute $\mathcal{Q}(\bm{\xi'};\bm{\xi})$ for each candidate $\bm{\xi'}$, and select the mode $\bm{\xi'}$ which maximises $\mathcal{Q}(\bm{\xi'};\bm{\xi})$.

\subsection{Extrapolating the modes to the surface}
\label{app:Extrapolation_method}

To extrapolate $\bm{\xi}$ to the surface,
the velocity of the selected $\bm{\xi'}$ is set such that its kinetic energy in the simulated region matches the one of $\bm{\xi}$.
By matching kinetic energies for both simulated and full models in the simulation region, this method assumes the balance of excitation and damping is the same in the whole star as in the truncated simulation domain.

\section{Empirical method for surface velocities}
\label{app:Belkacem_explanation}

To obtain the empirical surface velocities of the full stellar model using scaling laws we follow the approach in \citet{Bordadagua2025}, updated from \citet{Belkacem2015b}. The scaling law for the $\nu_\mathrm{max}'$ of the full stellar model is:

\begin{equation}
    \nu_\mathrm{max}' = \left(\frac{M}{M_\odot}\right)\left(\frac{T_\mathrm{eff}}{T_{\mathrm{eff},\odot}}\right)^{3.5}\left(\frac{L}{L_\odot}\right)\nu_{\mathrm{max},\odot}
    \label{eq:numax_compute}
\end{equation}

where any quantity indicated with $\odot$ represents the solar value, in particular we set $T_{\mathrm{eff},\odot} = 5777\ \mathrm{K}$, $\nu_{\mathrm{max},\odot} = 3050\ \mu\mathrm{Hz}$. The surface velocity amplitude at $\nu_\mathrm{max}'$, $V_{\mathrm{max},0}'$, is then computed using

\begin{equation}
    V_{\mathrm{max},0}' =  \zeta_0^{-1} \left(\frac{M}{M_\odot}\right)^{-0.89}\left(\frac{T_\mathrm{eff}}{T_{\mathrm{eff},\odot}}\right)^{-5.48}\left(\frac{L}{L_\odot}\right)^{0.36}V_{\mathrm{max},\odot},
\end{equation}

with $\zeta_0 = 0.59$ the non-adiabatic coefficient from \citet{Samadi2012}, and the maximum mode surface velocity of the Sun $V_{\mathrm{max},\odot} = 18.5$ $\mathrm{cm}\ \mathrm{s}^{-1}$\citep{Samadi2010}. The exponents in this scaling relation are based on the ones reported in \citet{Vrard2018}.

The surface velocities for the radial modes, $V'_0(\nu)$, are computed using a Gaussian centred on $\nu_\mathrm{max}'$, with FWHM set to $\delta\nu_\mathrm{env} = 70\ \mu\mathrm{Hz}$, according to
\begin{equation}
    V'^2_0(\nu) = V_{\mathrm{max},0}'^2\mathrm{e}^{-4(\nu-\nu_\mathrm{max}')^2/(\delta\nu_\mathrm{env}^2/\mathrm{ln} \; \! 2)}.
\end{equation}
Finally, the surface velocities of the non-radial modes, $V'_\ell(\nu)$, are computed by relating them to the surface velocities of the radial modes through the ratio of their modes masses using eq.~14 of \citet{Belkacem2015b}:
\begin{equation}
    \left(\frac{V'_\ell(\nu)}{V'_0(\nu)}\right)^2 \approx \left(\frac{\mathcal{M}'_0}{\mathcal{M}'_\ell}\right)^2,
\end{equation}
where $\mathcal{M}'_0$ and $\mathcal{M}'_\ell$ are the radial and non-radial mode masses respectively, defined as
\begin{equation}
    \mathcal{M'_\ell}\equiv\int |{\bm{\xi}'}|^2/|\bm{\xi}'(r_\star)|^2 \: \mathrm{d}m.
\end{equation}
The velocity throughout the domain of each mode of the full stellar model can then be obtained by normalising these modes such that their surface velocity matches the one computed using this empirical method.\\
The surface velocity of each non-radial mode is computed by considering the surface velocity of the radial mode with the closest frequency to that of the non-radial mode. All non-radial modes that are closest in frequency to a chosen radial mode therefore use the surface velocity of that radial mode in the computation. This can result in saw-tooth patterns because the surface velocity of the radial mode suddenly jumps from one non-radial mode to the next as the different frequency means a different radial mode is selected. We neglect the radiative damping in the interior, which would otherwise reduce the surface velocities of the non-radial modes \citep{Bordadagua2025}. The empirical surface velocities we report can therefore be thought of as an upper bound estimate.




\bibliography{references}{}
\bibliographystyle{aasjournalv7}



\end{document}